\documentclass[acmtog]{acmart}
\acmSubmissionID{613}

\usepackage[ruled]{algorithm2e} 

\SetAlFnt{\small}
\SetAlCapFnt{\small}
\SetAlCapNameFnt{\small}
\SetAlCapHSkip{0pt}

\DeclareUnicodeCharacter{2194}{\ensuremath{\leftrightarrow}}

\setcopyright{cc}
\setcctype{by-nc-nd}

\acmJournal{TOG}
\acmYear{2025}
\acmVolume{44}
\acmNumber{4}
\acmArticle{}
\acmMonth{8}
\acmPrice{}

\acmDOI{10.1145/3731173}

\usepackage{wrapfig}
\usepackage{pifont}
\usepackage[nameinlink,capitalise]{cleveref}
\usepackage{tabularx} 
\usepackage{wasysym}

\usepackage{booktabs} 

\usepackage[most]{tcolorbox}
\tcbset{on line, boxsep=0pt, left=1pt, right=1pt, top=1pt, bottom=1pt, colback=yellow, colframe=yellow, sharp corners}
\usepackage{soul} 

\DeclareGraphicsExtensions{.png,.jpg,.pdf,.ai,.psd}
\newcommand{\IG}[1]{}
\newcommand{\LL}[1]{}
\newcommand{\TL}[1]{}
\newcommand{\VDE}[1]{}
\newcommand{\YHG}[1]{}
\newcommand{\todo}[1]{}
\newcommand{\JY}[1]{}
\newcommand{\MH}[1]{}

\renewcommand{\IG}[1]{{\textcolor{orange}{[\textbf{IG:} #1]}}}
\renewcommand{\LL}[1]{{\textcolor{blue}{[\textbf{LL:} #1]}}}
\renewcommand{\TL}[1]{{\textcolor{red}{[\textbf{TL:} #1]}}}
\renewcommand{\VDE}[1]{{\textcolor{green}{[\textbf{VDE:} #1]}}}
\renewcommand{\YHG}[1]{{\textcolor{teal}{[\textbf{YHG:} #1]}}}
\renewcommand{\todo}[1]{{\textcolor{red}{[#1]}}}
\renewcommand{\JY}[1]{{\textcolor{orange}{[\textbf{JY:} #1]}}}
\renewcommand{\MH}[1]{{\textcolor{purple}{[\textbf{MH:} #1]}}}

\newcommand{\rgbi}{\textsf{RGB}\xspace}
\newcommand{\rgbx}{\textsf{RGB$\rightarrow$X}\xspace}
\newcommand{\xrgb}{\textsf{X$\rightarrow$RGB}\xspace}
\newcommand{\rgbxx}{\textsf{RGB$\leftrightarrow$X}\xspace}
\newcommand{\rgbxrgb}{\textsf{RGB$\rightarrow$X$\rightarrow$RGB}\xspace}

\newcommand{\loss}{\mathcal{L}}
\newcommand{\schedule}{\bar{\alpha}}

\newcommand{\eps}{\boldsymbol{\varepsilon}}
\newcommand{\model}{{\eps}_\theta}

\newcommand{\z}{\mathbf{z}}
\newcommand{\condi}{\mathbf{c}_\mathrm{i}}
\newcommand{\condiin}{\mathbf{c}_\mathrm{i+}}
\newcommand{\condiinedited}{\condiin^\text{edited}}
\newcommand{\condiout}{\mathbf{c}_\mathrm{i-}}
\newcommand{\condp}{\mathbf{c}_\mathrm{p}}
\newcommand{\nulli}{{\emptyset}_\mathrm{i}}
\newcommand{\nullp}{{\emptyset}_\mathrm{p}}

\newcommand{\One}{\raisebox{-0.4mm}{\fontsize{10.5}{9}\selectfont\ding{192}}\xspace}
\newcommand{\Two}{\raisebox{-0.4mm}{\fontsize{10.5}{9}\selectfont\ding{193}}\xspace}
\newcommand{\Three}{\raisebox{-0.4mm}{\fontsize{10.5}{9}\selectfont\ding{194}}\xspace}
\newcommand{\Four}{\raisebox{-0.4mm}{\fontsize{10.5}{9}\selectfont\ding{195}}\xspace}
\newcommand{\Five}{\raisebox{-0.4mm}{\fontsize{10.5}{9}\selectfont\ding{196}}\xspace}

\definecolor{MyGreen}{RGB}{0, 153, 74}
\definecolor{MyRed}{RGB}{216, 34, 42}
\definecolor{MyBlue}{RGB}{0, 151, 205}
\newcommand{\kept}{\textcolor{MyGreen}{\ding{52}}}
\newcommand{\edited}{\textcolor{MyBlue}{\ding{70}}}
\newcommand{\dropped}{\textcolor{MyRed}{\ding{56}}}

\title{IntrinsicEdit: Precise generative image manipulation in intrinsic space}

\author{Linjie Lyu}
\orcid{0009-0007-4763-8457}
\affiliation{%
    \institution{Max-Planck-Institute for Informatics, Saarland Informatics Campus}
    \city{Saarbr\"ucken}
    \country{Germany}
}
\affiliation{%
    \institution{Adobe Research}
    \city{London}
    \country{UK}
}
\email{llyu@mpi-inf.mpg.de}

\author{Valentin Deschaintre}
\orcid{0000-0002-6219-3747}
\affiliation{%
    \institution{Adobe Research}
    \city{London}
    \country{UK}
}
\email{deschain@adobe.com}

\author{Yannick Hold-Geoffroy}
\orcid{0000-0002-1060-6941}
\affiliation{%
    \institution{Adobe Research}
    \city{Quebec}
    \country{Canada}
}
\email{holdgeof@adobe.com}

\author{Milo\v{s} Ha\v{s}an}
\orcid{0000-0003-3808-6092}
\affiliation{%
    \institution{Adobe Research}
    \city{San Jose}
    \country{USA}
}
\email{mihasan@adobe.com}

\author{Jae Shin Yoon}
\orcid{0000-0003-0181-4869}
\affiliation{%
    \institution{Adobe Research}
    \city{San Jose}
    \country{USA}
}
\email{jaeyoon@adobe.com}

\author{Thomas Leimk\"uhler}
\orcid{0009-0006-7784-7957}
\affiliation{%
    \institution{Max-Planck-Institute for Informatics, Saarland Informatics Campus}
    \city{Saarbr\"ucken}
    \country{Germany}
}
\email{thomas.leimkuehler@mpi-inf.mpg.de}

\author{Christian Theobalt}
\orcid{0000-0001-6104-6625}
\affiliation{%
    \institution{Max-Planck-Institute for Informatics, Saarland Informatics Campus}
    \city{Saarbr\"ucken}
    \country{Germany}
}
\email{theobalt@mpi-inf.mpg.de}

\author{Iliyan Georgiev}
\orcid{0000-0002-9655-2138}
\affiliation{%
    \institution{Adobe Research}
    \city{London}
    \country{UK}
}
\email{igeorgiev@adobe.com}

\begin{document}


\begin{teaserfigure}
    \centering
    \includegraphics{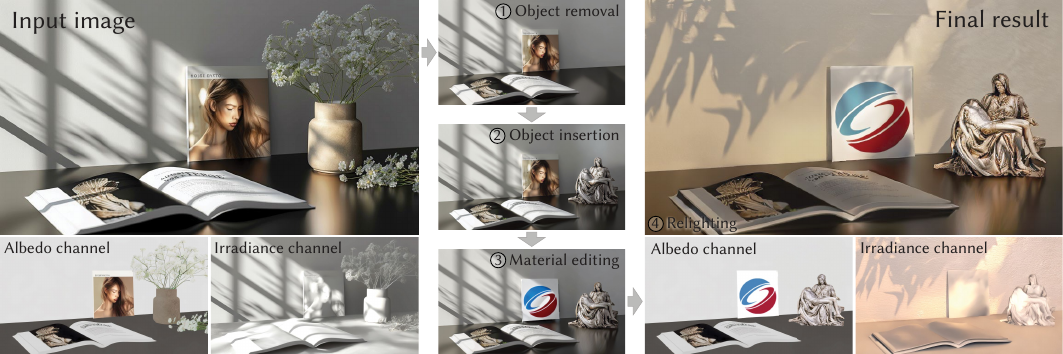}%
    \caption{
        We propose a generative framework for diverse image-editing tasks, where precise manipulations can be performed in an intrinsic-image space and global-illumination effects are subsequently resolved automatically. Here we show a progressive transformation of an input image: \One~We first remove the flowers and the vase from the albedo channel and then \Two~insert a new object in that channel. \Three~We replace the texture of another object before \Four~relighting the scene using a new irradiance channel. After each intrinsic-channel manipulation, we can render a physically plausible result. No single prior method can perform all these edits and provide similar levels of precision and identity preservation while delivering comparable image quality.\\
    }
    \label{fig:teaser}%
\end{teaserfigure}


\begin{abstract}
Generative diffusion models have advanced image editing by delivering high-quality results through intuitive interfaces such as prompts, scribbles, and semantic drawing. However, these interfaces lack precise control, and associated editing methods often specialize in a single task. We introduce a versatile workflow for a range of editing tasks which operates in an intrinsic-image latent space, enabling semantic, local manipulation with pixel precision while automatically handling effects like reflections and shadows. We build on the \rgbxx diffusion framework and address its key deficiencies: the lack of identity preservation and the need to update multiple channels to achieve plausible results. We propose an edit-friendly diffusion inversion and prompt-embedding optimization to enable precise and efficient editing of only the relevant channels. Our method achieves identity preservation and resolves global illumination, without requiring task-specific model fine-tuning. We demonstrate state-of-the-art performance across a variety of tasks on complex images, including material adjustments, object insertion and removal, global relighting, and their combinations.
\end{abstract}


%

%
%


\maketitle

\section{Introduction}

Image editing is a fundamental operation in the creative domain. Editing tasks range from subtle local corrections to more substantial modifications, such as altering the appearance and layout of objects or adjusting lighting. Achieving high-fidelity edits has traditionally required significant expertise and time. We propose a generative method that operates in an intrinsic-image space and significantly simplifies multiple non-trivial editing tasks: object insertion/removal, material manipulation, relighting.

Generative diffusion models~\cite{sohl2015deep,ho2020denoising,rombach2022high} have recently revolutionized imaging, as researchers realized that such models are capable of not only generating new images from text prompts, but can be repurposed for inpainting and other non-trivial edits. Recent work has introduced intuitive interfaces based on prompting~\cite{brooks2023instructpix2pix,sheynin2024emu}, dragging~\cite{shi2024dragdiffusion,wu2025draganything}, scribbling~\cite{ding2024training,lee2024scribble}, or semantic drawing~\cite{zhang2023adding,zhu2025champ}. Despite ease of use, such methods are typically limited to relatively high-level control, making it challenging to precisely define the desired edit while keeping the remaining image content intact. A partial solution to preserve identity is to mask the pixels that should remain unaffected~\cite{avrahami2022blended}. However, this can be difficult when intricate, non-local effects such as shadows, reflections, and color bleeding need to be considered, as these typically have fuzzy boundaries that are hard to anticipate.

\begin{figure}[t]
    \includegraphics{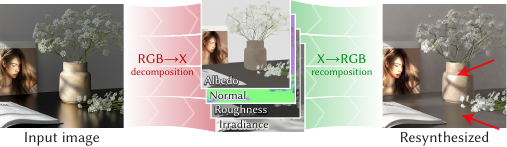}%
    \caption{
        \textbf{\rgbxx overview.}
        An \rgbx diffusion model decomposes a given image into intrinsic channels, while a complementary neural rendering \xrgb diffusion model composes channels into an image \cite{zeng2024rgb}. The complete image-to-image \rgbxrgb pipeline promises semantic editing with pixel precision by manipulating the channels before recomposition. Unfortunately, the models' generative nature causes random identity shifts in the resynthesized image, and successful editing requires adjusting multiple entangled channels, hindering usability. We address both these issues to unlock the image-editing potential of \rgbxx.
    }
    \label{fig:rgbx}
\end{figure}

Methods that perform intrinsic image decomposition and re-synthesis~\cite{luo2024intrinsicdiffusion,kocsis2024intrinsic,zeng2024rgb} promise accurate control over the entire image by representing it through channels that encode per-pixel information about geometry, appearance, and lighting. The \rgbxx framework of~\citet{zeng2024rgb}, illustrated in \cref{fig:rgbx}, envisions an editing pipeline where one (i)~decomposes the input image into intrinsic channels using an \rgbx model, (ii)~applies adjustments to these channels, and finally (iii)~recomposes an edited image using a neural rendering \xrgb model. This approach is inspired by classical 3D workflows where geometry, appearance, and lighting are defined and manipulated independently, enabling precise, physically based editing.

However, the practical realization of this promising vision requires solving the challenges of (i)~identity preservation and (ii)~the need to edit multiple channels simultaneously. Indeed, unlike a complete 3D-scene representation, the intrinsic-image space encodes only a subset of the information required to perfectly reproduce the input image, leaving room for the neural renderer to sample from an entire distribution of images consistent with the intrinsic conditions, necessarily shifting identity. Furthermore, the intrinsic channels carry redundant information; removing or inserting an object by (say) editing the albedo channel requires non-trivial updates to other channels to render a faithfully edited result.

In this work, we address both the identity preservation and channel entanglement limitations of the \rgbxx framework to  unlock its full image-editing potential. Furthermore, we do so exclusively with inference-time techniques, without requiring any further training. We first ensure that we can reconstruct the input image from the estimated intrinsic channels, which is necessary to preserve identity. We achieve this by performing exact inversion of the \xrgb model. The resulting noise vector may contain too much image-specific information baked in, which hinders editability. To avoid that, prior to inversion we optimize the originally unused \xrgb text-prompt embedding to absorb such information and encode it as a model condition. Second, to address the entanglement of intrinsic channels and gain the freedom to edit only the best-suited one(s) for the current task, we encode the remaining channels into the prompt embedding, at the same time as we optimize it for the aforementioned edit-friendly inversion. This channel-to-prompt transfer enables the model to preserve the non-edited properties of the input image more abstractly and flexibly than direct per-pixel specifications, while still allowing for precise, localized editing of the channel(s) of interest, making for a streamlined editing process.

Our technical advancements combine to enable a wide range of image editing tasks within a single framework. This framework features an interpretable latent space---the intrinsic images, which allows for pixel-level control through both traditional and modern image manipulation tools. Moreover, our method can achieve edits that other approaches struggle to perform well, such as pasting a texture onto an existing object, seamlessly integrating an inserted 3D object with specular reflections, or fully relighting a scene---as shown in \cref{fig:teaser}, all while automatically resolving global illumination effects. In summary, our main contributions are:
\begin{itemize}
    \item A diffusion inversion method targeting editability through intrinsic-channel conditions;
    \item Intrinsic-channel disentanglement for streamlined editing;
    \item Applications to diverse tasks, including appearance editing, object insertion and removal, and relighting of indoor scenes, showing automatic resolution of global illumination effects.
\end{itemize}

\section{Related work}

\paragraph{Image generation}

Over the past decade, generative models for image synthesis have gained significant attention. Early approaches focused on variational auto-encoders~\cite{kingma2013auto}, generative adversarial networks~\cite{goodfellow2014generative,karrasstylegan2019}, and normalizing flows~\cite{kobyzev2020normalizing}. More recently, diffusion models~\cite{sohl2015deep,ho2020denoising} have emerged as the state of the art, offering unprecedented image quality and diversity through training scalability~\cite{rombach2022high,ramesh2022hierarchical,peeblesdit2019,nichol2021glide,po2024state}. A central challenge to image authoring is control over the generation~\cite{zhang2023adding, bhat2024loosecontrol}.

\paragraph{Generative image editing}

Generative methods have quickly gained attention for the decades-old challenge of realistic image editing \cite{zhu2016generative,abdal2019image2stylegan,shen2020interpreting,collins2020editing,avrahami2022blended,sheynin2024emu}. Edits can be specified through various modalities. A common one is text, where the user describes the desired modification to an image through natural language~\cite{brooks2023instructpix2pix, sheynin2024emu, deutch2024turboedit}. While simple to use, text only supports high-level editing, making it challenging to obtain precise desired edits. Other approaches propose more localized modification, via cut and paste~\cite{alzayer2024magic} or dragging~\cite{endo2022user,pan2023drag,pandey2024diffusionhandles, mou2024dragondiffusion}, offering a more precise interface.

\paragraph{Diffusion inversion and prompt optimization}

DDIM inversion \cite{song2021denoising} approximately encodes the original image into the noise that initiates the DDIM diffusion sampling, and is widely used in editing tasks. Null-text inversion (NTI) \cite{mokady2023null} improves identity preservation by introducing pivot tuning for the null text embeddings, at the cost of more expensive inference-stage optimization. Negative-text inversion \cite{miyake2023negative} and the work of \citet{han2023improving} alleviate the computation cost of NTI, with a trade-off in identity preservation. Several works have proposed performance improvements for the noise inversion process \cite{garibi2024renoise,pan2023effective}. Edit-friendly DDPM \cite{huberman2024edit} is fast and can achieve accurate image reconstruction. However, it can over-entangle the inverted noise with the image, causing ghosting artifacts during editing (as we show in \cref{fig:ablation_inversion}). In our work, we adopt exact DDIM inversion \cite{hong2024exact}. Although it involves an optimization for each diffusion step, it maps the image to a compact, single initial noise and guarantees identity preservation and good editability. We further improve editability via a prompt-tuning stage \cite{gal2022image,kawar2023imagic,mahajan2024prompting,chung2023prompt} before inversion. Unlike previous prompt-tuning methods such as Imagic~\cite{kawar2023imagic}, ours encodes a broader range of intrinsic image information through additional supervision, resulting in more controllable edits.

\paragraph{Intrinsic decomposition and re-rendering}

Several recent works make progress in intrinsic image decomposition \cite{barrow1978recovering}, leveraging advances in diffusion models \cite{kocsis2024intrinsic,chen2025intrinsicanything,luo2024intrinsicdiffusion}. Our work builds atop the \rgbxx framework~\cite{zeng2024rgb} which uses intrinsic channels (albedo, normals, etc.) to control image generation and editing. However, a straightforward application of this framework suffers from loss of identity and from the need to make aligned edits to several channels at once. We tackle these limitations through noise inversion and prompt-embedding optimization, preserving the identity and naturally blending the edits without modifying the rest of the image.

\paragraph{Single-image relighting}

Relighting is a challenging task that requires reasoning about geometry, appearance, and lighting. Existing methods often focus on constrained scenarios, such as the relighting of portraits \cite{ponglertnapakorn2023difareli,zhang2024iclight,nestmeyer2020learning,sun2019single}, single objects \cite{jin2024neural,zeng2024dilightnet}, outdoor \cite{griffiths2022outcast} and indoor scenes \cite{Zhang2024Latent,murmann2019dataset,li2022physically}. These methods typically employ an explicit lighting model, encoded by an environment map or some parametric light-source representation. Inspired by classical intrinsic image decomposition \cite{barrow1978recovering}, recent approaches started employing shading (irradiance) maps as their lighting representation, notably for relighting outdoor scenes \cite{kocsis2024lightit,yu2020self} and compositing \cite{zhang2024zerocomp}. Such ``shading map'' representation offers several advantages. Compared to spherical (HDR) environment maps, shading maps have lower dynamic range and are the same size as the image, simplifying their ingestion into neural networks and enabling concatenation. Our method adopts this shading map---estimated by \rgbxx{}---as lighting representation and offers a more versatile editing framework than methods explicitly designed for relighting.


\paragraph{Object insertion/removal}

Adding or removing content is a staple in the image editing toolbox~\cite{fielding2013techniques,niu2021making}. While seamless blending~\cite{burt1983multiresolution,perez2003poisson,farbman2009poisson} and harmonization~\cite{sunkavalli2010multi,xue2012understanding,tsai2017deep} have been investigated for decades, recent work often fine-tunes diffusion models in a supervised manner~\cite{winter2024objectdrop, chen2023anydoor, zhang2023controlcom,liang2024photorealistic}, specifying the object to insert/remove through an image and/or mask. Unlike our method, these approaches are specialized to this task. Closest to our work is ZeroComp~\cite{zhang2024zerocomp} which proposes to composite the intrinsic channels of the foreground object and background before generating the edited version.

\paragraph{Material editing}

Materials are crucial to a scene's appearance. While their editing is trivial in 3D, it is difficult on a 2D image due to potentially complex global-illumination interactions. Recent work focuses on material editing on objects, either through sliders~\cite{sharma2024alchemist,delanoy2022generativematerials} or image exemplars~\cite{cheng2025zest}. Material editing at the scene level has been shown in \rgbxx but suffers from the identity drifts mentioned earlier.

\begin{figure*}%
    \includegraphics{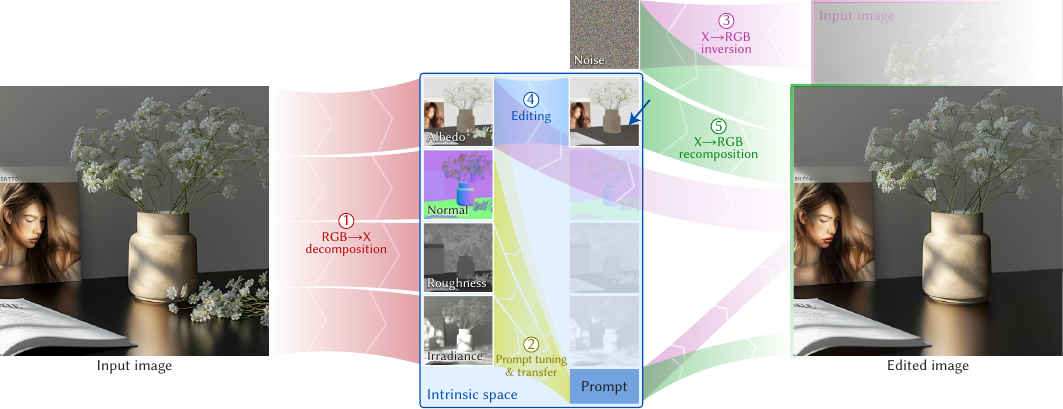}%
    \caption{
        \textbf{IntrinsicEdit overview.}
        We outline our intrinsic-space editing pipeline, here showing the removal of flowers by manipulating the albedo channel.
        \One~We run \rgbx to decompose the input image into intrinsic channels.
        \Two~We tune the prompt embedding to the image and channels (\cref{sec:prompt_tuning}). We also select a subset of channels for editing (here albedo only); any other channels that are entangled with that subset for the desired edit are transferred to the prompt and subsequently dropped (\cref{sec:channel_transfer}). This step allows us to edit a single channel while preserving information from the rest.
        \Three~We perform exact \xrgb inversion w.r.t.\ the remaining conditions, i.e.\ kept channels and optimized prompt. This step finds the noise map that, together with the conditions, accurately reconstructs the input image (\cref{sec:inversion}).
        \Four~We can now perform the desired edit by manipulating the selected channels.
        \Five~Finally, we feed those channels to the \xrgb model, along with the optimized prompt and inverted noise, to synthesize the edited image (\cref{sec:image_editing}).
        This pipeline allows us to alter only certain image modalities (e.g.\ material), while automatically propagating the changes to a realistic result and preserving untouched aspects.
    }
    \label{fig:overview}
\end{figure*}

\section{Method}
\label{sec:method}

Our goal is to leverage the strong natural-image priors of large generative models for realistic image manipulation on a variety of tasks, such as object insertion/removal, material editing, and relighting. We use the intrinsic-image ``latent space'' of the \rgbxrgb diffusion pipeline~\cite{zeng2024rgb} which (i)~decomposes an image into intrinsic channels (albedo, normal, roughness, irradiance) and (ii)~recomposes them after edits. However, this pipeline has critical limitations that we must address to make it practicable.

\paragraph{Identity shift}

The immediate challenge we face stems from the generative nature of the \xrgb rendering model. Appending that model to the \rgbx decomposition introduces randomness in the middle of the pipeline. In a generative setting, that randomness is necessary as it enables sampling from the entire distribution of images consistent with a given set of intrinsic channels. For our image-editing application that randomness is harmful: it causes identity drifts in the output \rgbi image---even without edits! To avoid loss of identity, we need to anchor the \xrgb model to reproduce the input image from the initial intrinsic decomposition. This is an inversion problem---computing the starting noise for \xrgb inference that yields a target output image for given (intrinsic) conditions. We perform exact diffusion inversion~\cite{hong2024exact} on \xrgb to find that noise (\cref{sec:inversion}). Once found, we fix it; editing then involves manipulating the intrinsic channels and resynthesizing by running \xrgb using that same noise.

\paragraph{Image/noise entanglement}

The ability to accurately reconstruct the input image from the initial intrinsic channels does not necessarily imply that manipulating those channels will produce a plausible edited result. In particular, we observed that the inversion often ``bakes'' a lot of image-specific information into the noise, which leads to unrealistic editing results containing artifacts or ghost features from the input image. We attribute this to the inverted noise being outside the Gaussian-distribution ``comfort zone'' of the \xrgb model, (i)~as a result of weak or inaccurate conditioning from the channels (e.g.\ unsupported materials like fabrics), (ii)~due to their imprecise \rgbx estimation, or (iii)~due the input image being outside both models' distributions. As a remedy, to bring the noise closer to its expected distribution and improve editability, prior to inversion we tune the (originally unused) \xrgb text-embedding conditioning to encode image information that the inversion would otherwise bake into the noise (\cref{sec:prompt_tuning}).

\paragraph{Inter-channel entanglement}

Being equipped with the reconstruction noise and necessary conditions, we can start manipulating intrinsic channels to achieve our desired edit. This is where we hit our third problem, which is a general limitation of intrinsic channels: they are partially entangled with one another. For example, removing an object requires careful, aligned editing of \emph{all} channels! While inpainting the albedo channel may be simple, plausibly adjusting the irradiance demands an infeasible light simulation. The blessing of providing dense, semantic conditioning comes with the curse of having to edit all pixels in sync; otherwise, conflicts among channels will lead to artifacts in the \xrgb output (see \cref{fig:ablation_prompt}, bottom row).

Our solution is simple: we drop any channels that conflict with the desired edit on our chosen channel(s). However, inversion w.r.t.\ a reduced number of conditions can lead to the aforementioned noise-baking problem. We apply a similar remedy: we optimize the prompt to take over the conditioning from the channels that are to be dropped. This optimization effectively transforms the pixel-precise intrinsic conditioning to a more abstract one (\cref{sec:image_editing}). While this solution significantly compresses the amount of conditioning information, it maintains editability and makes the entire pipeline practical by providing freedom to select the most suitable editing modality while delivering plausible results in our experiments.

\paragraph{Overview}

\Cref{fig:overview} illustrates our editing pipeline. After obtaining an \rgbx intrinsic decomposition of the input image, we optimize the \xrgb prompt embedding to (i)~tune it to the image and intrinsic conditions and (ii)~take over conditioning from edit-entangled channels which are subsequently dropped. We then invert \xrgb w.r.t.\ the kept channel(s) and optimized prompt, to obtain a reconstruction noise map. The map remains fixed throughout the editing which involves manipulating the kept channel(s) and rendering out a result by invoking \xrgb. Next, we describe these steps in detail.

\subsection{\xrgb inversion}
\label{sec:inversion}

\xrgb is a latent diffusion model~\cite{rombach2022high} that reverses an iterative process
\begin{equation}
    \label{eq:diffusion}
    \z_t = \sqrt{\schedule_t} \z_0 + \sqrt{1 - \schedule_t} \eps
\end{equation}
which gradually corrupts the latent-space representation $\z_0$ of an image with Gaussian noise $\eps \sim \mathcal{N}(\mathbf{0},\mathbf{1})$. The noising schedule $\schedule_t \in [0,1]$ is a function of the time step $t \in [0,T]$, such that ($\schedule_0, \schedule_T) = (1, 0)$ and thus $\z_T \sim \mathcal{N}(\mathbf{0},\mathbf{1})$. The model implements a neural network $\model$, with parameters $\theta$, which predicts the noise in a given noisy latent $\z_t$ and which is trained with the objective
\begin{equation}
    \label{eq:ddpm_loss}
    \loss(\theta) = \mathbb{E}_{\z_0, t, \eps}\,\Big\| \, \eps - \model\big(\z_t, t, \condi,\condp \big)\Big\|^2.
\end{equation}
The model is conditioned on a text-prompt embedding $\condp$ and a set $\condi$ of (latent representations of) intrinsic channels---albedo, normal, roughness, irradiance. We omit metallicity which we found unreliable, as also pointed out by \citet{zeng2024rgb}. Note also that they parameterize the model to predict velocity instead of noise~\cite{salimans2022progressive}.

Given a set of conditions, we can apply deterministic (DDIM) sampling~\cite{song2021denoising} to iteratively transform an initial noise sample $\z_T \sim \mathcal{N}(\mathbf{0},\mathbf{1})$ to a clean (latent) image $\z_0$ via the recurrence
\begin{equation}
    \label{eq:ddim}
    \z_{t-1} = \sqrt{\alpha_{t-1}\alpha_t^{-1}} \z_t + \left(\!\sqrt{\alpha_{t-1}^{-1} \! - 1}-\sqrt{\alpha_t^{-1} \! - 1}\right) \cdot \model\big(\z_t, t, \condi,\condp\big).
\end{equation}
We can think of this $T$-step sampling process as a neural photo-realistic renderer that generates an image given a Gaussian-noise sample, a set of intrinsic channels, and a prompt embedding:
\begin{equation}
    \label{eq:xrgb}
    \z_0 = \text{\xrgb}\big(\z_T, \condi, \condp \big).
\end{equation}

To be able to reconstruct our input image without any edits, we need to invert the above rendering function to obtain the noise $\z_T$ that reproduces the image's clean latent $\z_0$ given conditions $\condp$, $\condi$. We use exact DDIM inversion \citep{hong2024exact} which we found to work significantly better than faster alternatives such as naive DDIM inversion or edit-friendly DDPM inversion~\cite{huberman2024edit}. Given $\z_0$, $\condi$, and $\condp$, the inversion performs gradient-descent optimization of the trajectory of latents $\{\z_t\}_{t=1}^T$ using the following recurrence, starting from $t=1$:
\begin{equation}
    \label{eq:exact_inversion}
    \z_t = \operatorname*{arg\,min}_{\z'_t} \Big\| \z_{t-1} - \underbrace{\z'_{t-1}(\z_t, t, \condi,\condp)}_\text{\cref{eq:ddim}} \Big\|^2,
\end{equation}
taking $\z_{t-1}$ obtained in the previous step. That is, it finds the point $\z_t$ that is mapped to $\z_{t-1}$ by the DDIM sampling in~\cref{eq:ddim}.

\subsection{Prompt tuning}
\label{sec:prompt_tuning}

Exact inversion allows us to accurately reconstruct the input image but bakes into the optimized noise any image-identity information that is not contained in the conditions. This becomes particularly problematic for editing tasks, where much can simply not be altered because it is ``fixed'' by that noise. As a result, the \xrgb model can yield artifacts after intrinsic editing. To avoid over-entangling the noise with the image, \emph{prior to inversion} we tune the otherwise unused prompt embedding $\condp$ to pick up image-identity features absent from the intrinsic conditions $\condi$ using the following loss:
\begin{equation}
    \label{eq:prompt_tuning}
    \loss_\text{tune}(\condp) = \mathbb{E}_{t,\eps}\left\|\eps - \model\left(\z_t, t, \condi,\condp \right)\right\|^2.
\end{equation}
This is the same as the training loss in \cref{eq:ddpm_loss}, this time optimizing for the prompt $\condp$ with frozen model parameters $\theta$, given intrinsic conditions $\condi$ and input image $\z_0$. At each optimization iteration, we sample a time step $t$ uniformly, and noise $\eps \sim \mathcal{N}(\mathbf{0},\mathbf{1})$, and compute the latent $\z_t$ using \cref{eq:diffusion}. This prompt tuning pushes $\condp$ to contain as much residual information as possible about the input image without over-fitting to a single initial noise $z_T$.

It is also possible to optimize the estimated intrinsic conditions using the same loss, to align them better with the image and further mitigate feature noise baking. Unfortunately, such naive optimization makes them uninterpretable and thus uneditable.

\subsection{Channel-to-prompt transfer}
\label{sec:channel_transfer}

Having a tuned prompt and correspondingly inverted noise allows us to start editing our image though the intrinsic channels. For example, small object-color adjustments are achievable by manipulating only the albedo channel. However, more substantial color edits require carefully updating indirect lighting effects in irradiance, and removing or inserting objects requires non-trivial updates to \emph{all} channels, including shading and shadows in irradiance---a task arguably even less practical than directly manipulating the input image.

We address this problem as follows. For the given editing task, we first identify a subset of channels most suitable for direct manipulation, then drop any other entangled channels that would require manual adjustment. For example, adding/removing objects impacts all channels while albedo editing may not impact normals. Consequently, however, inversion with fewer condition could reintroduce the noise entanglement discussed in \cref{sec:prompt_tuning}. We apply a similar solution: we transfer the information from the dropped channels to the prompt embedding. We achieve this by optimizing the embedding $\condp$ such that \xrgb generation with the kept channels and the prompt yields a similar result to generation with all initial channels $\condi = \{\condiin, \condiout\}$ (kept $\condiin$ and dropped $\condiout$) and initial null prompt $\nullp$. The prompt-optimization loss is thus
\begin{equation}
    \label{eq:prompt_transfer}
    \begin{split}
        \loss_\text{transfer}(\condp) = \mathbb{E}_{t, \eps}\Big\|&\model\big(z_t, t, \{\condiin, \condiout\},\nullp \big)
        \,- \\[-1mm]
        &\model\big( z_t, t, \{\condiin, \nulli\}, \condp \big) \Big\|^2,
    \end{split}
\end{equation}
where $\nulli$ means using null intrinsic conditions in place of the dropped channels $\condiout$. Note that \cref{eq:prompt_transfer} requires the model $\model$ to support any combination $\{\condiin, \nulli\}$ of valid and null conditions. Luckily, \xrgb does as it was trained with channel dropout~\cite{zeng2024rgb}.

Intuitively, this optimization aims to make the prompt have the same effect as using the dropped intrinsic conditions but without requiring their per-pixel editing. The prompt $\condp$ now contains an abstract representation of all (dropped) conditions that are to be preserved, while the intrinsics $\condiin$ explicitly represent conditions to be edited, at the same time minimizing image-specific baking into the inverted noise. This disentangled representation allows us to edit the image in various ways while preserving its original identity.

In practice we perform a single prompt optimization, with a combined loss
\begin{equation}
    \label{eq:prompt_optimization}
    \loss_\text{prompt}(\condp) = \loss_\text{tune}(\condp) \, + \, \lambda \, \loss_\text{transfer}(\condp),
\end{equation}
where $\lambda$ balances between tuning and transfer; we use $\lambda \in [0.1,10]$ in our experiments.

\subsection{Intrinsic editing and final synthesis}
\label{sec:image_editing}

Given the kept intrinsic conditions $\condiin$, the optimized prompt $\condp$, and the noise $\z_T$ inverted for them, we can finally edit the intrinsics. The set of kept intrinsics varies per application, as we will detail in \cref{sec:results} below. Finally, we resynthesize an edited image using the edited intrinsics $\condiinedited$:
\begin{equation}
    \label{eq:image_edit}
    \z_0^\text{edited} = \text{\xrgb}\big(\z_T, \{\condiinedited, \nulli\}, \condp \big).
\end{equation}
We find that this approach leads to good image identity preservation while enabling precise and better disentangled manipulation of different image modalities.

Note that the inversion anchors \xrgb to the input \rgbi image and its corresponding \rgbx intrinsic decomposition, making the internals of our pipeline deterministic (except for the inherent randomness of stochastic inversion and prompt optimization). The full pipeline remains probabilistic: It takes a (random) noise input at the \rgbx entry point, and the edited result may vary depending on the estimated intrinsic channels as they are propagated to \xrgb.

\paragraph{Diffusion guidance.}

Guidance is widely used in diffusion models to boost generation quality. For \xrgb inversion, we do not apply guidance. During inference after editing, instead of classifier-free guidance (CFG)~\cite{ho2022classifier}, in \cref{eq:ddim} we replace $\model$ by
\begin{equation}
    \label{eq:guidance}
    \model^\text{guided} = \omega\,\model\big( \z_t, t, {\condi}^\text{edited},\condp \big) + (1-\omega)\, \model\big( \z_t, t, \condi,\condp \big).
\end{equation}
This form of guidance is similar to that in negative-prompt inversion~\cite{miyake2023negative} but uses the initial intrinsic condition $\condi$ instead of a null condition. We use guidance scale $\omega = 1.5$, following \citet{zeng2024rgb} who use the same scale for CFG in \xrgb.

\begin{figure*}[h!]%
    \includegraphics{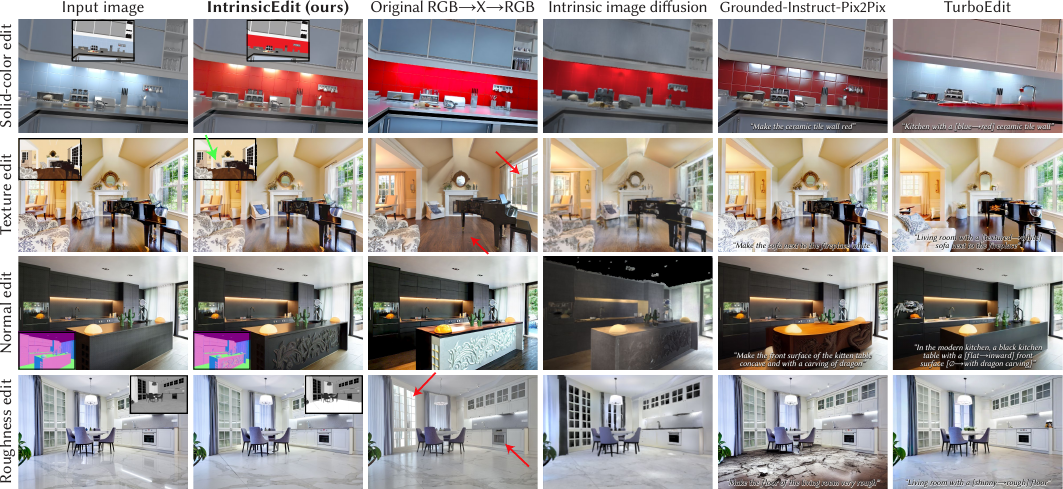}%
    \caption{
        \textbf{Material editing.}
        We compare our method against two intrinsic-image methods: original \rgbxrgb \cite{zeng2024rgb} and intrinsic image diffusion \cite{kocsis2024intrinsic}, and two prompt-based methods: Grounded-Instruct-Pix2Pix \cite{Groundedip2p} and TurboEdit \cite{deutch2024turboedit}. Prompt-based methods fail to provide fine-grained control. Ours is the only one that allows for precise manipulation of individual material properties, preserving identity and harmonizing the edits much better than prior intrinsic-space approaches. Notice the red wall in the top row matching the original material properties while correctly adjusting the color, including the reflection on the counter. The second row shows texture editing on the armchair and addition of two pillows, preserving the lighting and scene identity. In the third row, our approach automatically extends the wooden floor and preserves the kitchen island color despite editing only the normal map. The bottom row shows roughness editing, making the floor more matte and adjusting the reflections.
    }
    \label{fig:material_editing}
\end{figure*}

\begin{figure*}[h!]%
    \includegraphics{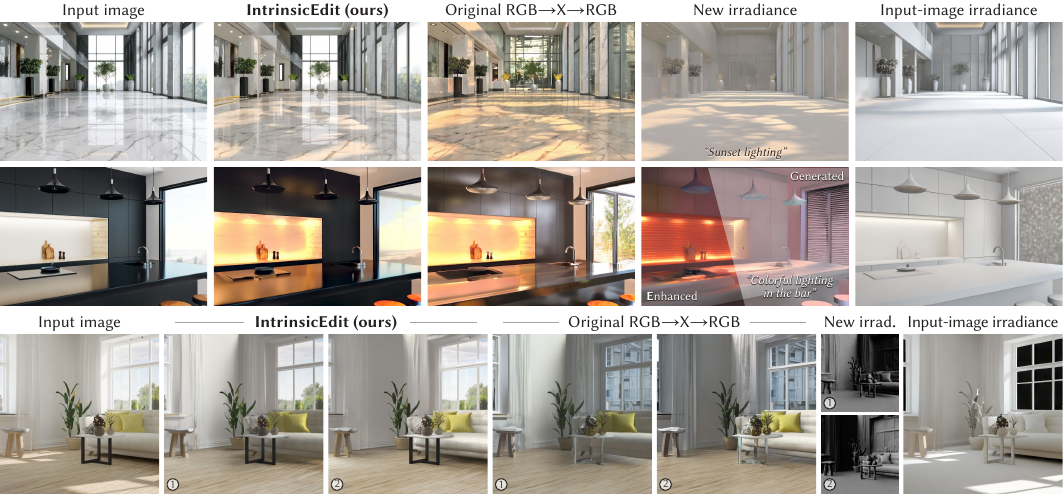}%
    \caption{
        \textbf{Relighting.}
        In the top two rows we generate a new irradiance channel via prompting as described in \cref{sec:relighting}. In the bottom row we generate novel irradiance maps using the volumetric shading model of OutCast \cite{griffiths2022outcast}. Our method handles the new lighting condition more naturally than original \rgbxrgb relighting \cite{zeng2024rgb}, even when the change is drastic (second row), and better preserves the identity of the scene content.
    }
    \label{fig:relighting}
\end{figure*}

\section{Results}
\label{sec:results}

We now present an evaluation of our approach on four applications: material editing, object removal, object insertion, and relighting. Additionally, we show quantitative evaluations for a subset of applications, as well as ablations to validate our inversion and prompt optimization strategies. Our supplemental document contains an expanded set of results.

\paragraph{Implementation details}

We implemented our method atop the public code and models of \citet{zeng2024rgb}. We run our prompt optimization with the loss in \cref{eq:prompt_optimization} for 200 iterations using AdamW optimizer~\cite{adamw} and learning rate 0.1. To invert the 50-step \xrgb inference w.r.t.\ the optimized prompt, we follow the backward Euler DDIM solver of~\citet{hong2024exact}, performing 2-3 optimization iterations per diffusion step, seeded by naive DDIM inversion. For roughness editing, we use guidance strength $\omega = 6$ and prompt-loss balance $\lambda \in [1,10]$ ($\lambda = 1$ in \cref{fig:material_editing}) to ameliorate the \xrgb model's weakness and enforce the edit.

The original \rgbxrgb pipeline of \citet{zeng2024rgb} is a baseline in all our results. For that baseline we use their \emph{inpainting} \xrgb model for all applications except relighting, as we found it to work significantly better than their base \xrgb model in that pipeline, especially for object removal and insertion. The inpainting model takes the input image, a box mask, and the intrinsic channels, and renders an output only inside the mask. We compute the mask by doubling the bounding box of the \emph{edit mask} (difference between original and edited channel(s)) along each dimension, to accommodate for illumination effects; substantial edits can yield masks that cover the entire image. We always use the base (non-inpainting) \xrgb model in our pipeline.

\paragraph{Test data}

Most images in our qualitative evaluation are obtained from a stock image database. A small subset is from the Hypersim~\cite{roberts2021hypersim}, MIT Indoor~\cite{2009Recognizing}, and InteriorVerse~\cite{zhu2022learning} synthetic datasets, and we captured a few images ourselves using smartphone. Note that we did not specifically aim to collect synthetic-looking stock images (known synthetic is top row in \cref{fig:material_editing}, from HyperSim), but did focus on indoor imagery. We do not possess intrinsic-decomposition or edit ground truths for any of the images in our qualitative evaluation.

For real-world quantitative evaluation we use an object-removal dataset of 12 image pairs produced by taking pictures before and after manual object placement. We also evaluate material editing on a dataset derived from 10 synthetic 3D scenes, produced by changing an object's albedo or roughness, and rendering 14 before/after pairs.

\paragraph{Channel organization}

For each edit we show, we specify which channels are kept~(\kept), kept and edited~(\edited), or dropped~(\dropped) in the inline table below. Any dropped channels are transferred to the prompt%
\setlength{\columnsep}{4mm}
\begin{wrapfigure}{r}{0.56\columnwidth}
    \vspace{-2mm}
    \setlength{\tabcolsep}{1.5pt}
    \scalebox{0.8}{
        \begin{tabularx}{0.69\columnwidth}{lcccc}
            \toprule
            \textbf{Edit} &  \textbf{Albedo} & \textbf{Normal}  & \textbf{Roughn.} & \textbf{Irrad.}\\
            \midrule
            Color & \edited & \kept & \kept & \dropped \\
            Normal & \dropped & \edited & \dropped & \dropped \\
            Roughness\!\! & \kept & \kept / \dropped & \edited & \dropped \\
            Relighting\! & \kept & \kept & \kept & \edited \\
            Removal & \edited & \dropped & \dropped & \dropped \\
            Insertion & \edited & \edited / \dropped & \dropped & \dropped \\
            \bottomrule
        \end{tabularx}
    }
    \vspace{-2mm}
\end{wrapfigure}
as per \cref{{sec:channel_transfer}}. For relighting, we found that transferring the unedited irradiance to the prompt improves the inverted-noise disentanglement and the plausibility of edited results. For normal editing, we need to drop the albedo because geometry can change significantly; our supplemental document shows drastic normal edits. For our quantitative evaluation of roughness editing, we drop the normal whose \rgbx estimation is unstable on specular surfaces. We do object insertion via only albedo or via both albedo and normal.

For original \rgbxrgb we follow the same channel organization, but we do not transfer dropped channels to the prompt.

\begin{figure*}[h!]%
    \includegraphics{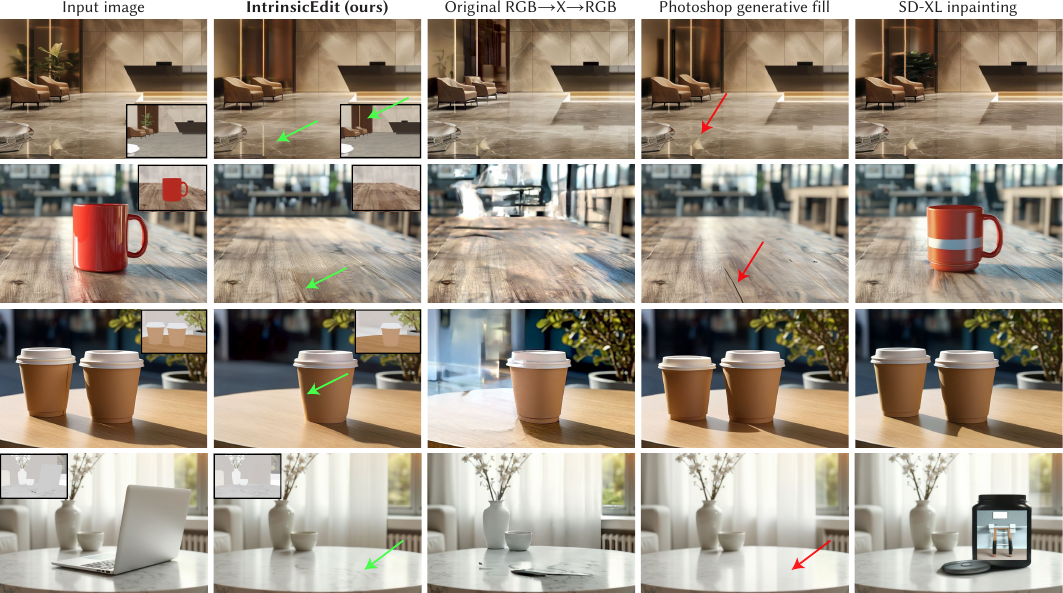}%
    \caption{
        \textbf{Object removal.}
        We compare against original \rgbxrgb \cite{zeng2024rgb}, Photoshop generative fill \cite{photoshop}, and Stable Diffusion XL inpainting \cite{sdxl-inpainting}. Without being specialized for this task, our method performs on par with or better than prior work. In the top row, notice the correct removal of the plant reflection from the marble floor. In the second and fourth rows, the table's texture is preserved better. In the third row, our method successfully removes the left cup, including the shadow it casts on the other cup, while previous methods even struggle to remove the cup completely.
    }
    \label{fig:removal}
\end{figure*}

\subsection{Qualitative evaluation}
\label{sec:results_qualitative}

\paragraph{Material editing}

Material editing targets the modification of surface color (texture), normal, or roughness. Such editing is greatly simplified in our framework where these intrinsic properties are directly available for manipulation. \Cref{fig:material_editing} shows that our method provides fine-grained editing control and generates results that harmonize better with the surrounding environment. We observe that it handles reflection of edited surfaces well (top two rows), and our normal editing correctly infers the right material for the modified kitchen island (third row). Finally, making a shiny floor matte achieves a realistic result that preserves the lighting in the scene.

The only other method with results not too far off from ours is the original \rgbxrgb~\cite{zeng2024rgb}. Unfortunately it suffers from all the issues we have addressed in this paper. \citeauthor{zeng2024rgb} had to rely on precise masking to achieve \emph{some} identity preservation, at the cost of compromising the propagation of global illumination effects that the \xrgb model is otherwise able to deliver.

\paragraph{Relighting}
\label{sec:relighting}

Since the irradiance channel can be challenging to modify manually, we use a bootstrapping approach: We sample \xrgb with all intrinsic conditions except irradiance, using text description of the lighting, until we obtain the desired effect. While this approach is already a form of relighting, it leads to identity shifts. To use the new lighting with our method, we simply decompose the obtained image using \rgbx and use the extracted irradiance channel to relight the original input image using our pipeline, after potential stylistic manipulations to the channel.

We show relighting results in \cref{fig:relighting} where we change the orientation and color tint of incoming light, e.g.\ giving an impression of shadows from plants (first row) or turning on kitchen spotlights (second row). We also use the shading model of OutCast \cite{griffiths2022outcast} for relighting (third row), where scene depth is estimated, projected into a volumetric model, and queried for novel light directions.

\paragraph{Object removal}

We can remove an object from a photograph by inpainting the albedo channel using a photo editor's remove tool. We show results on multiple images in \cref{fig:removal}. We can see in all cases that the objects' reflections and shadows are well removed, and that the inpainted regions preserve the background identity better than previous work (second and fourth rows). Image-space inpainting methods have the disadvantage of requiring larger masks that enclose effects like shadows and reflections, which leads to identity shifts in large background regions. Our channel-inpainting masks can tightly bound the object, allowing our method to preserve the surroundings' identity. Note that each Photoshop generative-fill result is the subjectively best one selected from 5 samples.

\paragraph{Object insertion}

\begin{figure*}%
    \includegraphics{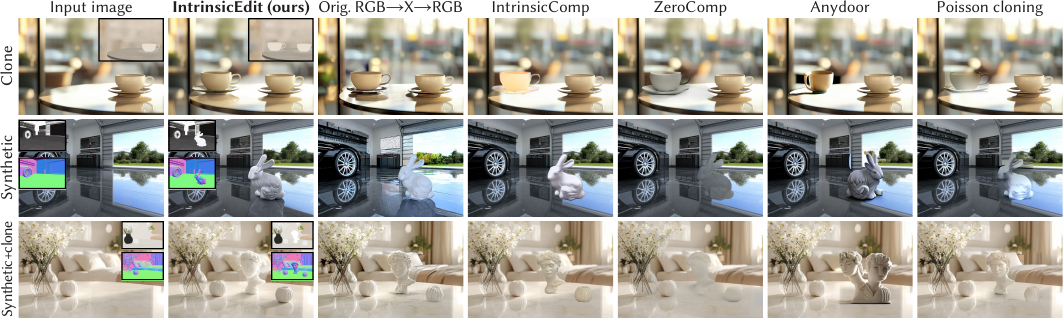}%
    \caption{
        \textbf{Object insertion.}
        We compare against original \rgbxrgb \cite{zeng2024rgb} and existing object-insertion and intrinsic-based methods: IntrinsicComp \cite{careaga2023intrinsic}, ZeroComp \cite{zhang2024zerocomp}, Anydoor \cite{chen2023anydoor}, and Poisson cloning \cite{perez2003poisson}. For intrinsic-based methods, we insert the object into the albedo channel (top row) or in both albedo and normal (middle and bottom rows). Despite not being specialized for this task, our approach better harmonizes the inserted object with the rest of the scene. This is particularly visible with the strong directional lighting in the bottom row with the added bust and sphere. In the top row, we duplicate an existing object, observing that our approach successfully handles the table reflection.
    }
    \label{fig:insertion}
\end{figure*}

We perform object insertion by decomposing an object image using \rgbx, or directly extracting its material channels if it is a synthetic object, and pasting them in the target image's channels. Similarly to object removal, we encode the irradiance information in the prompt. We show results in \cref{fig:insertion}, inserting objects by controlling the albedo map (top row), or both albedo and normal (middle \& bottom rows). Our method is the only one that can faithfully harmonize both the object and the rest of the scene, handling reflections and matching the lighting, with little to no compromise in identity.

\subsection{Quantitative evaluation}
\label{sec:results_quantitative}

We evaluate our method quantitatively on synthetic and real-world datasets with ground-truth before/after images. Editing starts from the ``before'' image and consumes no other data. For each dataset, we present visual comparisons and report PSNR and LPIPS \cite{zhang2018unreasonable} metrics, averaged over the dataset, w.r.t.\ the ``after'' ground truth for our method and the same baselines used in \cref{sec:results_qualitative}.

\paragraph{Synthetic material editing}

We evaluate color and roughness editing on renders of synthetic scenes, with dataset sizes of 10 and 4 respectively. \Cref{fig:material_editing_synthetic} summarizes the results. Although not perfect, our method produces images much closer to the reference edited results than all other methods, both visually and numerically.

\paragraph{Real-world object removal}

\Cref{fig:removal_real} summarizes our results on a real-world object-removal dataset, where we evaluate the results numerically over the whole image and only within the Photoshop generative-fill inpainting mask. Over the whole image, our method is close second behind Photoshop which achieves perfect pixel-value preservation outside its mask; our results suffer from diffusion latent-space encoding/decoding inconsistencies (see \cref{sec:ablation,fig:decoder_inversion}). Ours performs best within the Photoshop mask which necessarily has to be larger than our albedo-inpainting mask, to include any shadows and reflections. Note that masking can, in principle, be applied to our method, too.

\begin{figure}[ht]
    \includegraphics{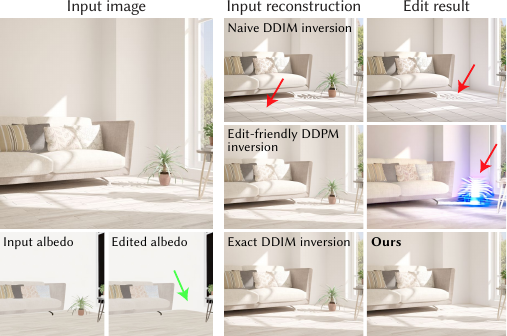}%
    \caption{
        \textbf{Inversion method ablation.}
        Replacing the exact DDIM inversion~\cite{hong2024exact} in our pipeline (\cref{fig:overview}) with naive DDIM~\cite{song2021denoising} or edit-friendly DDPM~\cite{huberman2024edit} inversion has a disastrous effect on its performance, here on object removal.
    }
    \label{fig:ablation_inversion}%
\end{figure}

\subsection{Ablation studies}
\label{sec:ablation}

\paragraph{Inversion method}

\Cref{fig:ablation_inversion} demonstrates that exact DDIM inversion~\cite{hong2024exact} is crucial to our method's performance and identity preservation (\cref{sec:inversion}). Replacing the inversion algorithm in our pipeline by edit-friendly DDPM inversion~\cite{huberman2024edit} preserves its ability to reconstruct the input image, but bakes too much information in its residual noise term, causing severe artifacts. Naive DIM inversion shows good editability but with loss of identity.

\paragraph{Prompt optimization}

\begin{figure}
    \includegraphics{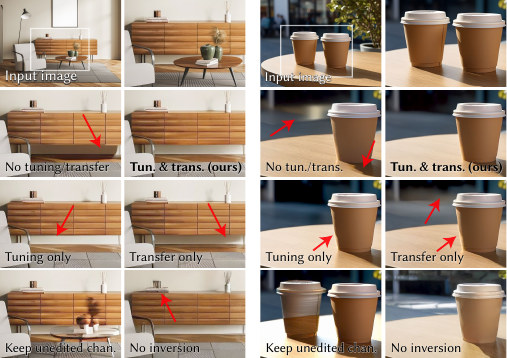}%
    \caption{
        \textbf{Prompt optimization ablation.}
        Excluding either prompt tuning or channel transfer from our method has negative consequences. Using tuning alone, the method struggles to accurately estimate the correct illumination. On the other hand, using transfer alone compromises identity preservation. Keeping the unedited channels instead of transferring them produces ghosting due to wrong geometry/illumination hints, while skipping inversion and running \xrgb inference with random noise leads to identity loss.
    }%
    \label{fig:ablation_prompt}%
\end{figure}

\Cref{fig:ablation_prompt} studies the impact of our prompt tuning and transfer optimizations (\cref{sec:prompt_tuning,sec:channel_transfer}). Without tuning or transfer, our method fails to preserve identity and handle illumination after object removal (wrong shadow, background shift). With tuning only, it struggles to estimate lighting (under cabinet) or fails to remove shadows (on remaining cup). With transfer only, it suffers from identity loss (background behind cups). Without inversion, sampling random noise for \xrgb inference, the prompt optimization can still encode part of the identity but cannot reproduce details in the input. Finally, if we keep the unedited channels as explicit conditions instead of transferring them to the prompt, they cause ghosting artifacts after object removal due to the wrong geometry hint. Our prompt transfer allows us to preserve information from entangled channels without having to edit them.

\section{Discussion}

\paragraph{Identity preservation}

\begin{figure}[t]
    \includegraphics{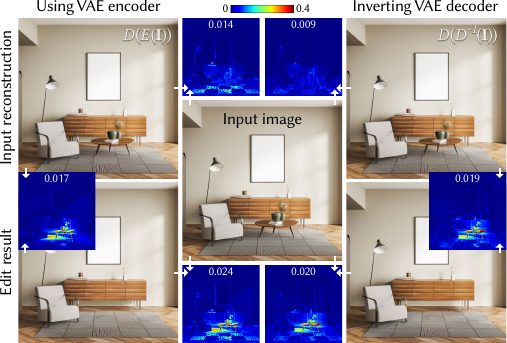}%
    \caption{
        \textbf{Identity preservation.}
        The variational autoencoder (VAE) of the base foundational (Stable Diffusion) model is a major source of   identify shifts in our method. Simply encoding and decoding the input image, $D(E(\mathbf{I}))$, leads to significant loss of fine detail. Inverting the decoder instead, $D(D^{-1}(\mathbf{I}))$, ameliorates the issue but does not eliminate it. Compared to those reconstructions, our edits produce mostly local and predictable differences. These are clear in the $L_1$ error maps (numbers are image averages).
    }
    \label{fig:decoder_inversion}
\end{figure}

Identity preservation, while significantly improved, is still not perfect. We found that the vast majority of identity shifts are caused by the latent-space encoding of the base Stable Diffusion~(SD) 2.1 model \cite{stablediffusion21} of \rgbxx. We explore the issue in \cref{fig:decoder_inversion}, where we compare two ways to obtain the latent representation of the input image: (1)~using the SD encoder and (2)~inverting the SD decoder \cite{hong2024exact}. Foregoing any editing, the former option (top left) shows significant loss in high-frequency detail after reconstruction (using the decoder). The latter (top right) yields a more accurate reconstruction, though not perfect---an indication of irrecoverable compression information loss. The error maps comparing our edited results to those reconstructions reveal that the image differences caused by our edits are mostly localized and predictable, accommodating the edit and its effect on the illumination.

\paragraph{Editing accuracy}

\begin{figure}[t]
    \includegraphics{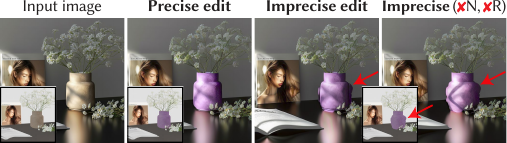}%
    \caption{
        \textbf{Editing accuracy.}
        Successful editing requires performing the necessary channel manipulation with accuracy; imprecision can lead to artifacts. Here, going outside the object's boundary with the albedo edit introduces a conflict with the normal-channel condition. Dropping the normal (and roughness) yields a plausible result but with altered object shape.
    }
    \label{fig:imprecise_edit}
\end{figure}

One of our main goals is to provide pixel-precise editing. However, achieving the expected result requires the user to perform the necessary channel manipulation with accuracy. In \cref{fig:imprecise_edit} we show the effect of an imprecise albedo manipulation on a color-editing task: editing the albedo beyond the object's boundary introduces a conflict with the geometry condition in the normal channel and yields artifacts. Dropping the normal produces a realistic result but alters the object's shape. Resolving such conflicts and ambiguities is an interesting challenge in generative image editing.

\paragraph{Channel inpainting}

Our object-removal pipeline still requires inpainting but on the intrinsic channels (e.g.\ albedo). In principle, this is a much easier task than inpainting the final image, and future work could investigate inpainting models specific to intrinsic channels to make this step more convenient and robust.

\paragraph{\rgbxx limitations}

Although our method shows high-quality results, it inherits some limitations from the \rgbxx models which are trained with limited indoor scene data; further evolution of these models will automatically improve our results. Roughness editing is less reliable than albedo, and metals and transparent objects remain challenging. Another limitation comes from occasionally imperfect \rgbx intrinsic-image decomposition, forcing our inversion process to bake more information into the noise, potentially limiting editing possibilities. More complex materials and light transport effects are another interesting future direction. For example, handling mirrors far from the edited object, multiple reflections, or editing materials such as skin, hair, fur, or fabrics remain challenging. Generally, precise illumination control in images remains an open problem, and is especially difficult for indoor scenes.

\begin{figure}[t]%
    \includegraphics{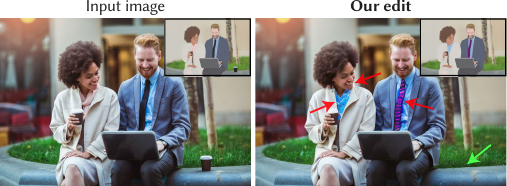}%
    \caption{
        \textbf{\rgbxx limitations.}
        Images outside the \rgbxx distribution remain challenging. While we can successfully remove the coffee cup, the background is uneditable due to poor intrinsic decomposition. People and garments remain difficult to edit realistically.
    }
    \label{fig:limitations}
\end{figure}

While we did not observe quality differences between synthetic- and real-looking images, our model performs best on data closer to the \rgbxx training distribution---indoor scenes; expanding to outdoor scenes and human characters will be key to future improvements as these currently pose a challenge. We show one such out-of-distribution example in \cref{fig:limitations}. While we can perform some successful editing on that image, the background is uneditable due to poor intrinsic decomposition, and people and their garments are generally not handled well, showing limited editing success.

\begin{figure*}[t]%
    \includegraphics{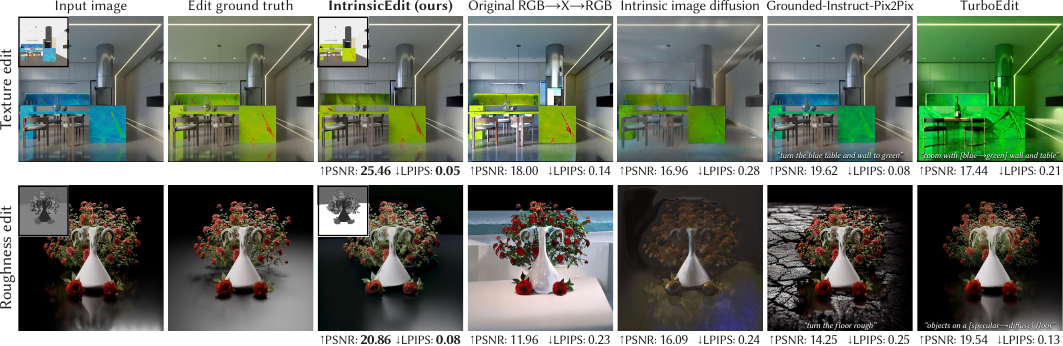}%
    \caption{
        \textbf{Material editing on synthetic images.}
        On a set of synthetic-scene renders, we compare our method against the same baselines as in \cref{fig:material_editing}, this time also quantitatively w.r.t.\ ground-truth edit results. The results of our method match the ground truths most closely, both numerically and visually, although not perfectly. The reported metrics are averaged over a dataset of 10 before/after pairs for texture editing and 4 pairs for roughness editing. We stress that we do not use any ground-truth intrinsic channels: editing starts from the ``before'' image and consumes no other data.
    }
    \label{fig:material_editing_synthetic}
\end{figure*}

\begin{figure*}[t]%
    \includegraphics{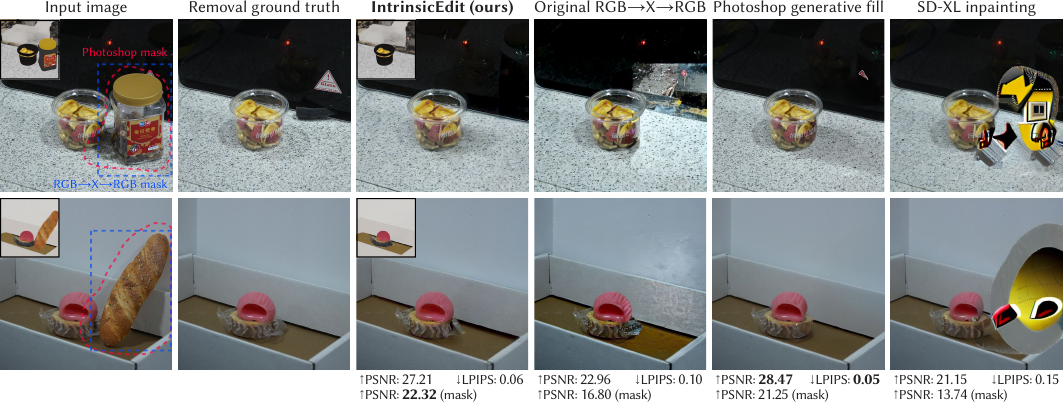}%
    \caption{
        \textbf{Real-world object removal.}
        We compare against the same baselines as in \cref{fig:removal}, this time also quantitatively w.r.t.\ ground-truth edit results. Photoshop generative fill is the only method that rivals ours, performing best over the entire image thanks to perfect pixel preservation outside its inpainting mask. Inside that mask, our method preserves identity best, textures in particular. The reported metrics are averaged over the dataset of 12 before/after pairs.
    }
    \label{fig:removal_real}
\end{figure*}

\paragraph{Speed and resolution}

The inference speed of our approach does not match that of feed-forward pipelines. Even though exact inversion is crucial for identity preservation, it can be slow. The resolution of our test images ranges from 512$\times$512 to 1920$\times$1080. For a 512$\times$512 image on an Nvidia H100 GPU, our method takes 75\,sec pre-editing (20\,sec for 4-channel \rgbx decomposition, 15\,sec for 200-step prompt optimization, 40\,sec for 50-step inversion) and 5\,sec for the 50-step \xrgb inference after editing. Processing a 1920$\times$1080 image takes approximately 500\,sec, mostly due to high memory demand for inversion. Large resolutions, such as 4K, are currently infeasible due to memory limitations. A future few-step (distilled) \xrgb model would greatly accelerate inversion as well as prompt optimization (due to smaller range of diffusion steps to sample). We can also reasonably expect future optimizations and hardware improvements to make our method more interactive.

\section{Conclusion}

Our approach provides pixel-precise generative image editing in intrinsic space. It is made possible by \xrgb diffusion inversion, optimizing both the noise and the prompt embedding to flexibly preserve non-edited information from the input image. We show the capabilities of our framework on a diverse set of non-trivial image manipulation tasks, including object insertion and removal, material editing, and full scene relighting. Despite some remaining limitations, our results demonstrate substantial progress towards unlocking the potential of generative decompose-edit-recompose approaches for image editing.


\begin{acks}

We obtained the 3D scenes for our synthetic material-editing dataset (\cref{fig:material_editing_synthetic}) from \href{https://www.blenderkit.com/}{Blenderkit}. Tianyu Wang and Qing Liu kindly provided the real-world object-removal dataset (\cref{fig:removal_real}). We thank Zheng Zeng for discussions and help with \rgbxx code and models.

\paragraph{Ethics}

We note that realistic generative image manipulation methods like ours could facilitate the spread of misinformation.

\end{acks}


\bibliographystyle{ACM-Reference-Format}
\bibliography{references}



\end{document}



\begin{teaserfigure}%
    \vspace{5mm}%
    \includegraphics{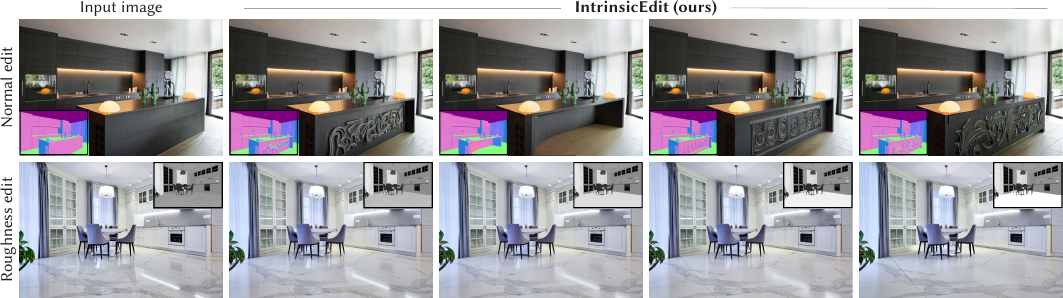}%
    \vspace{-1mm}
    \caption{
        \textbf{Normal and roughness editing.}
        We demonstrate several variants of material editing using our method. The top row shows different normal edits; note that we drop the albedo channel due to the conflicts with the changing geometry. Our method still delivers a plausible appearance on the floor. In the bottom row, we gradually increase the floor roughness from left to right.
    }
    \vspace{5mm}%
\end{teaserfigure}

\maketitle


In this supplemental document, we show qualitative results on material editing, object removal and insertion, and relighting, expanding the set presented in the main paper. We also include additional examples from our quantitative evaluation.

\vspace*{4mm}


\begin{figure*}%
    \includegraphics{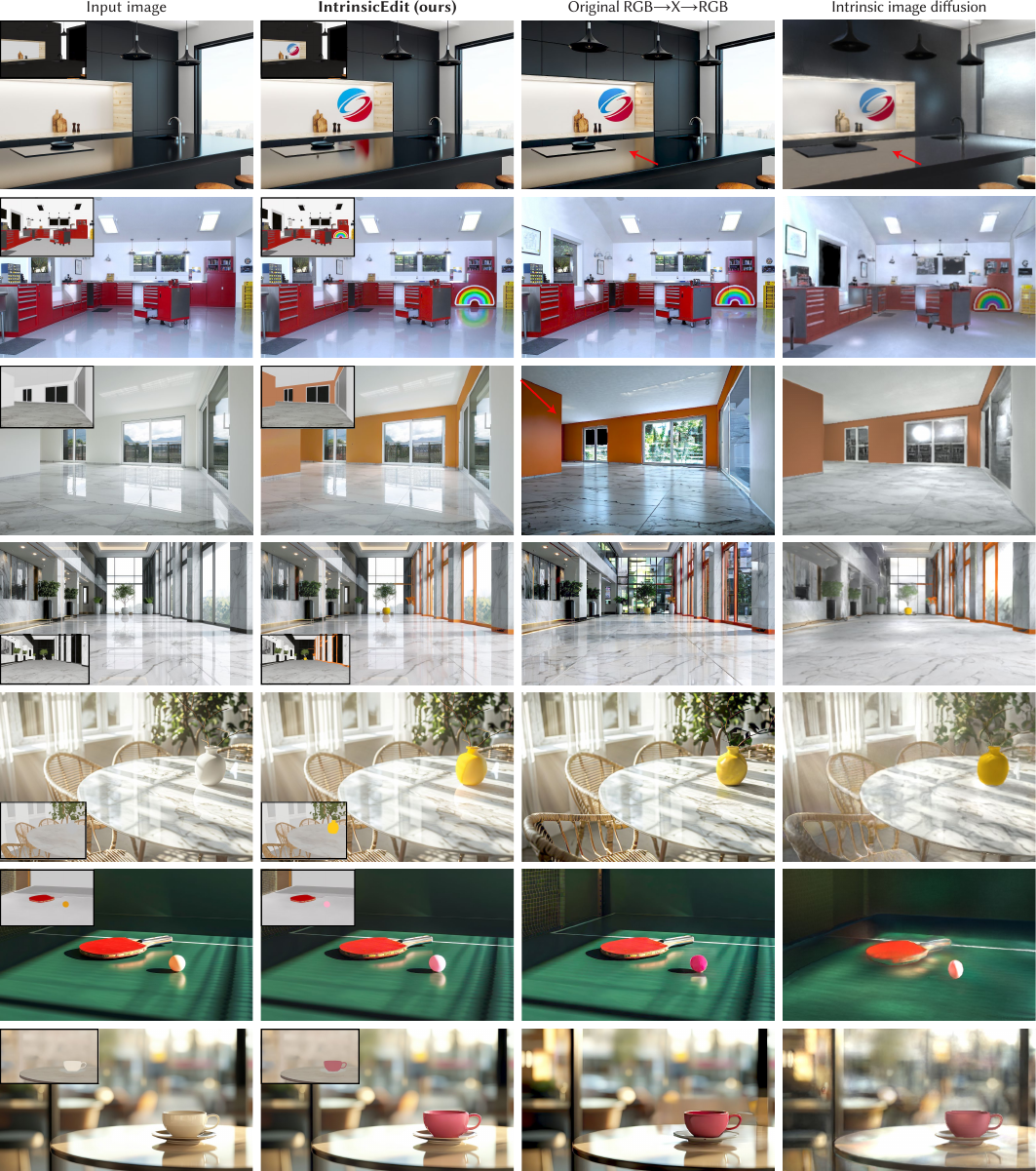}%
    \vspace{-2mm}
    \caption{
        \textbf{Color editing A.}
        We compare against two intrinsic-image methods: original \rgbxrgb \cite{zeng2024rgb} and intrinsic image diffusion \cite{kocsis2024intrinsic}. Our method enables precise manipulation of individual material properties while preserving identity and achieving seamless illumination harmonization.
    }%
\end{figure*}


\begin{figure*}%
    \includegraphics{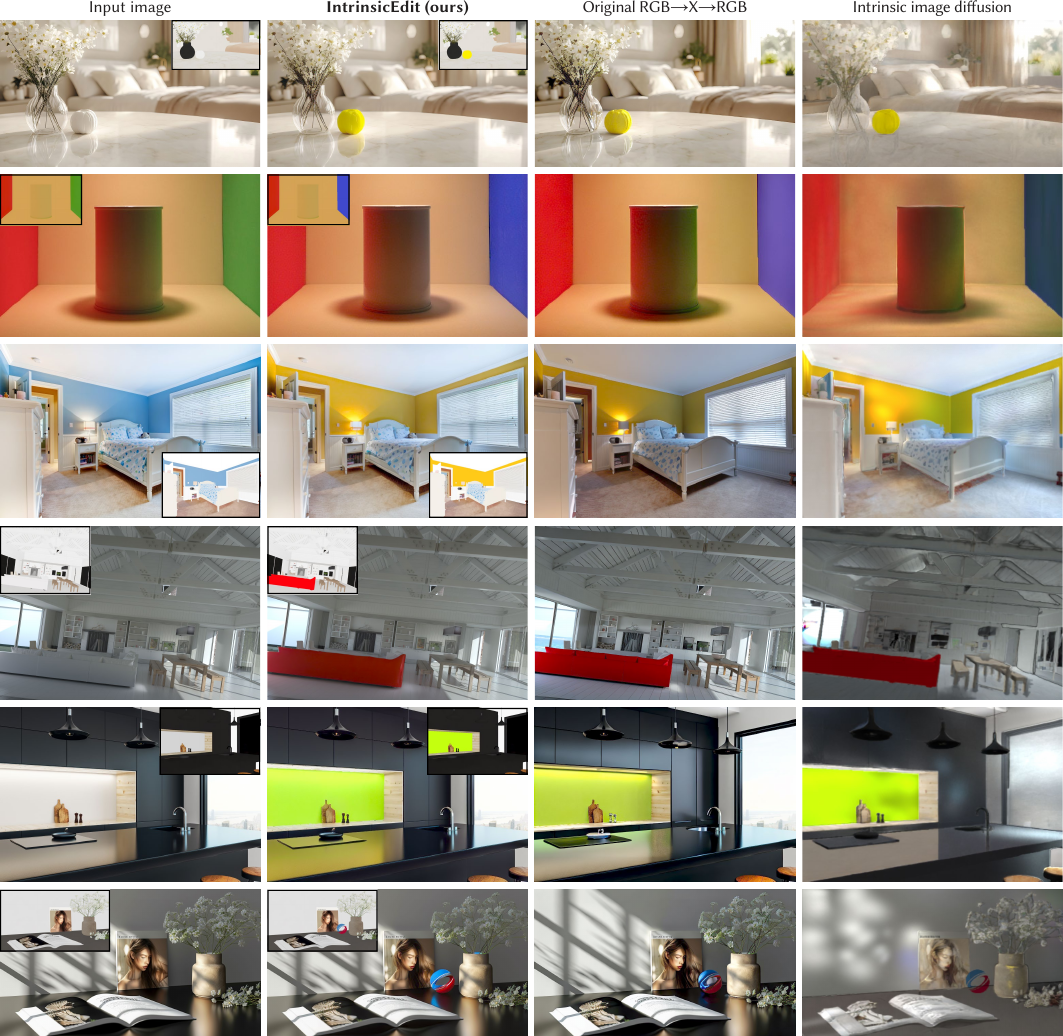}%
    \vspace{-2mm}
    \caption{
        \textbf{Color editing B.}
        We compare against two intrinsic-image methods: original \rgbxrgb \cite{zeng2024rgb} and intrinsic image diffusion \cite{kocsis2024intrinsic}. Our method enables precise manipulation of individual material properties while preserving identity and achieving seamless illumination harmonization. In the botom row, we observe that if the texture edit does not align with an existing object, our model interprets the edit as object insertion and harmonizes accordingly.
    }
\end{figure*}


\begin{figure*}%
    \includegraphics{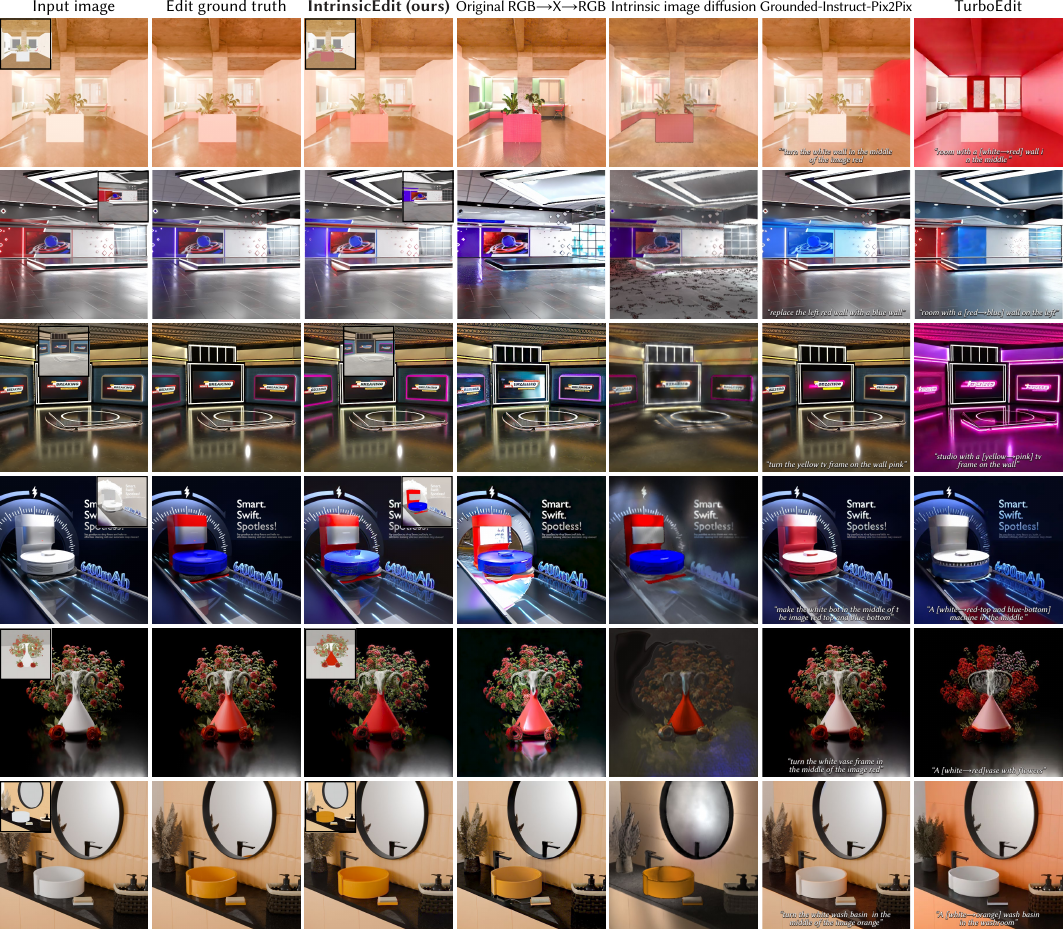}%
    \vspace{-2mm}
    \caption{
        \textbf{Synthetic color editing.}
        We include additional results used in the quantitative evaluation on a synthetic dataset in the paper.
    }
\end{figure*}


\begin{figure*}%
    \includegraphics{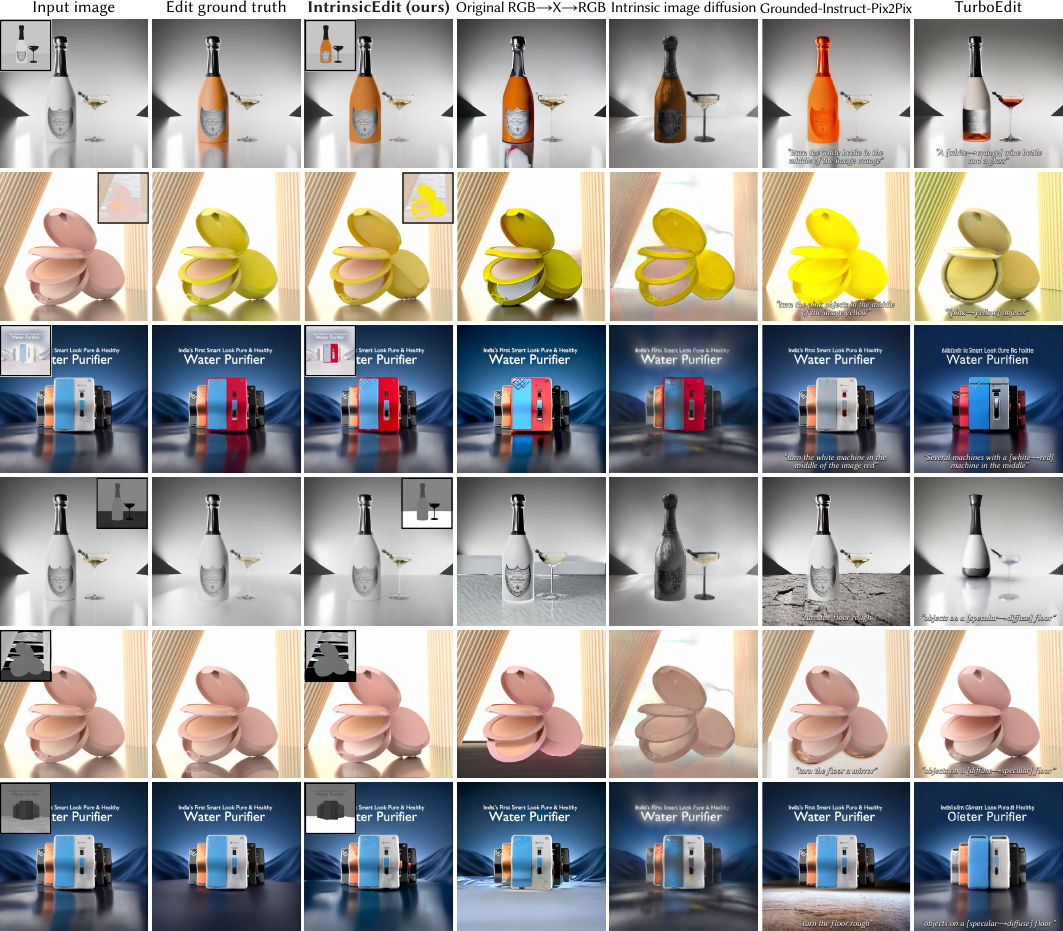}%
    \vspace{-2mm}
    \caption{
        \textbf{Synthetic color and roughness editing.}
        We include additional results used in the quantitative evaluation on a synthetic dataset in the paper. The top three rows show color editing, the bottom three rows show roughness editing.
    }
\end{figure*}


\begin{figure*}%
    \includegraphics{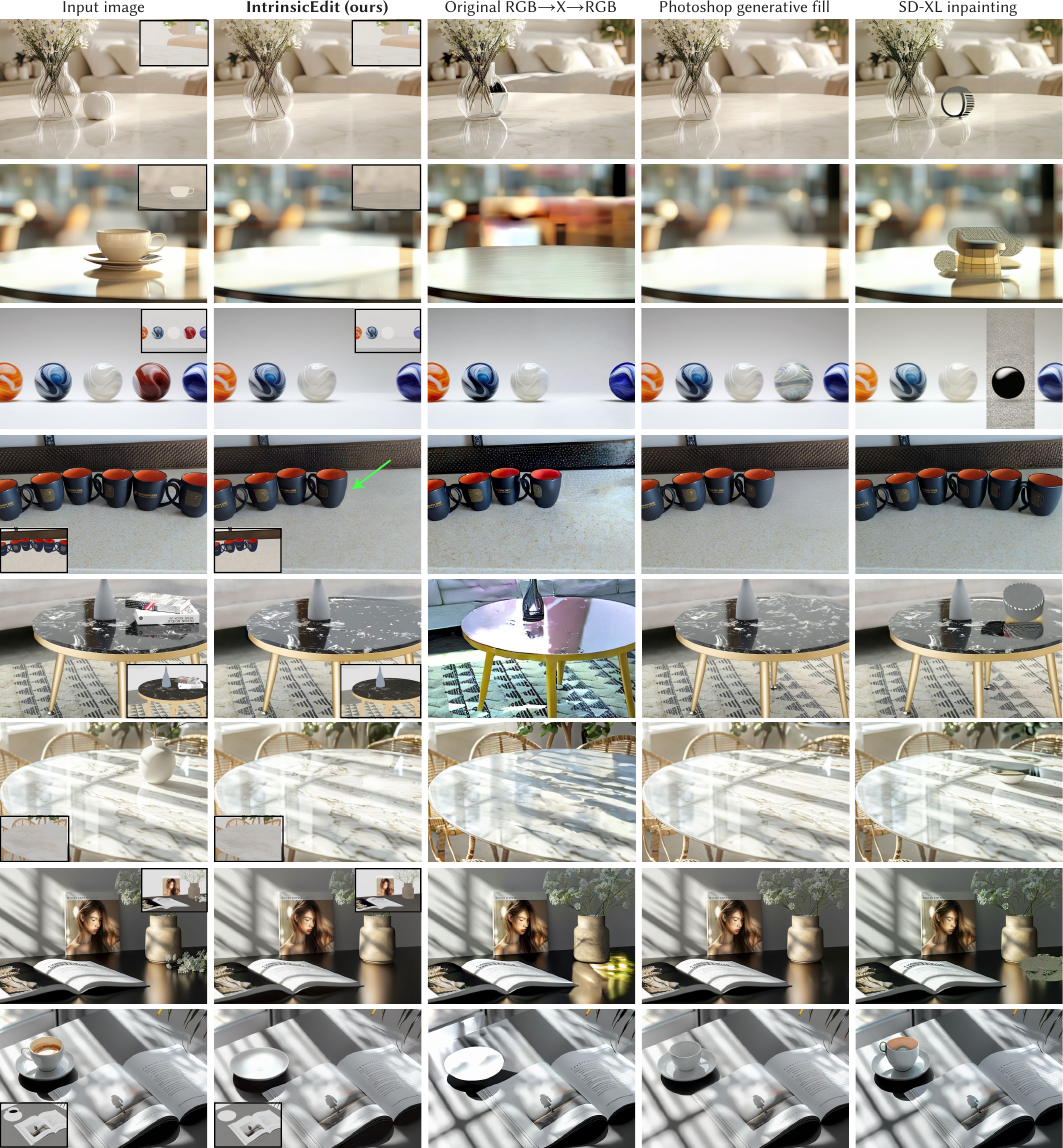}%
    \vspace{-2mm}
    \caption{
        \textbf{Object removal A.}
        We compare against original \rgbxrgb \cite{zeng2024rgb}, Photoshop generative fill \cite{photoshop}, and Stable Diffusion XL inpainting \cite{sdxl-inpainting}. Without being specialized for this task, our method performs on par with or better than prior work. It demonstrates a capability to automatically remove shadows and reflections in alignment with the object. Notably, in the third row, it successfully removes the shadow cast on the nearby cup.
    }
\end{figure*}


\begin{figure*}%
    \includegraphics{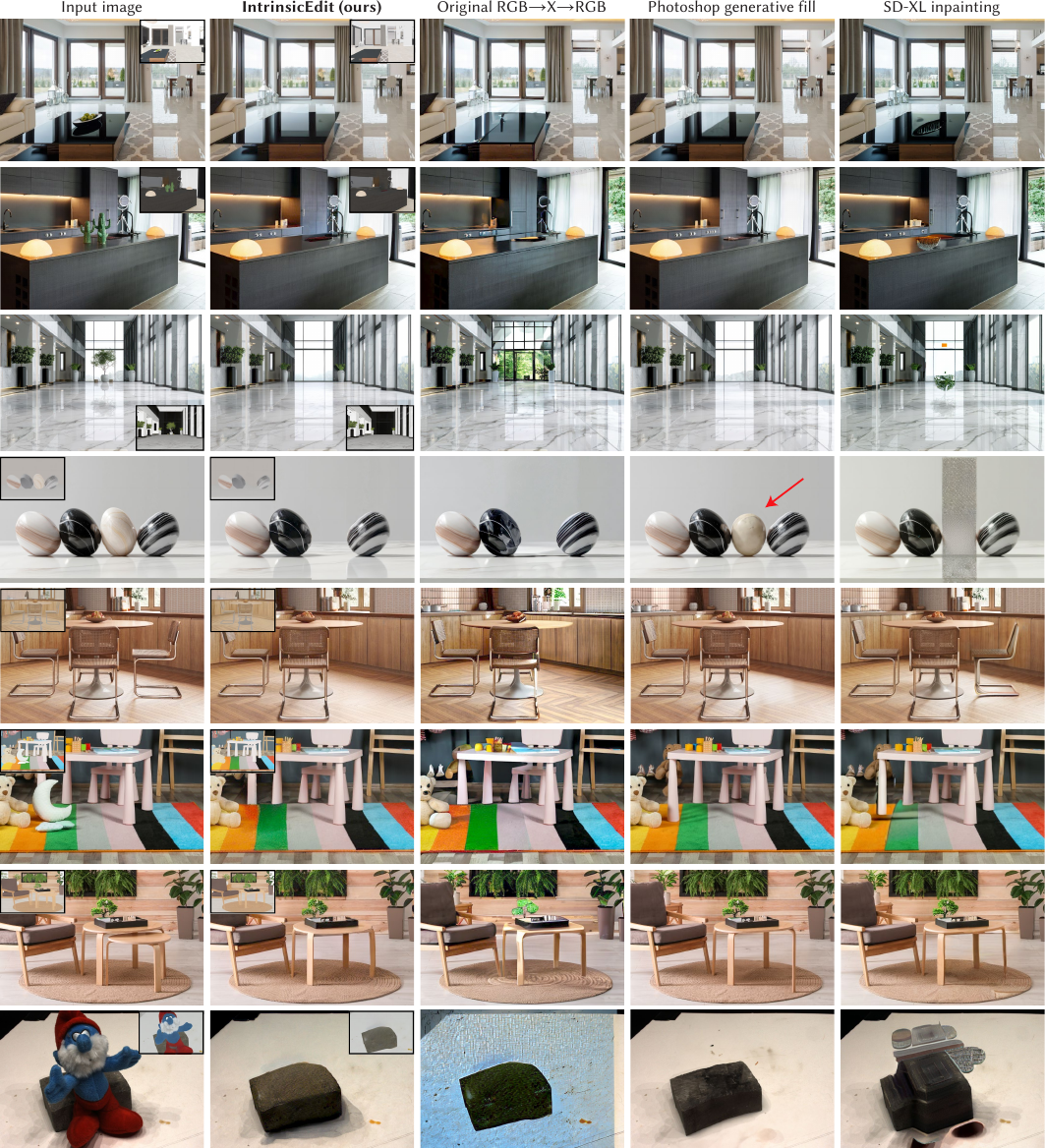}%
    \vspace{-2mm}
    \caption{
        \textbf{Object removal B.}
        We compare against original \rgbxrgb \cite{zeng2024rgb}, Photoshop generative fill \cite{photoshop}, and Stable Diffusion XL inpainting \cite{sdxl-inpainting}. Without being specialized for this task, our method performs on par with or better than prior work. It demonstrates a capability to automatically remove shadows and reflections in alignment with the object. Particularly, in the challenging case in the last row, our method still resolves multiple shadows after removal while previous methods fail.
    }
\end{figure*}


\begin{figure*}%
    \includegraphics{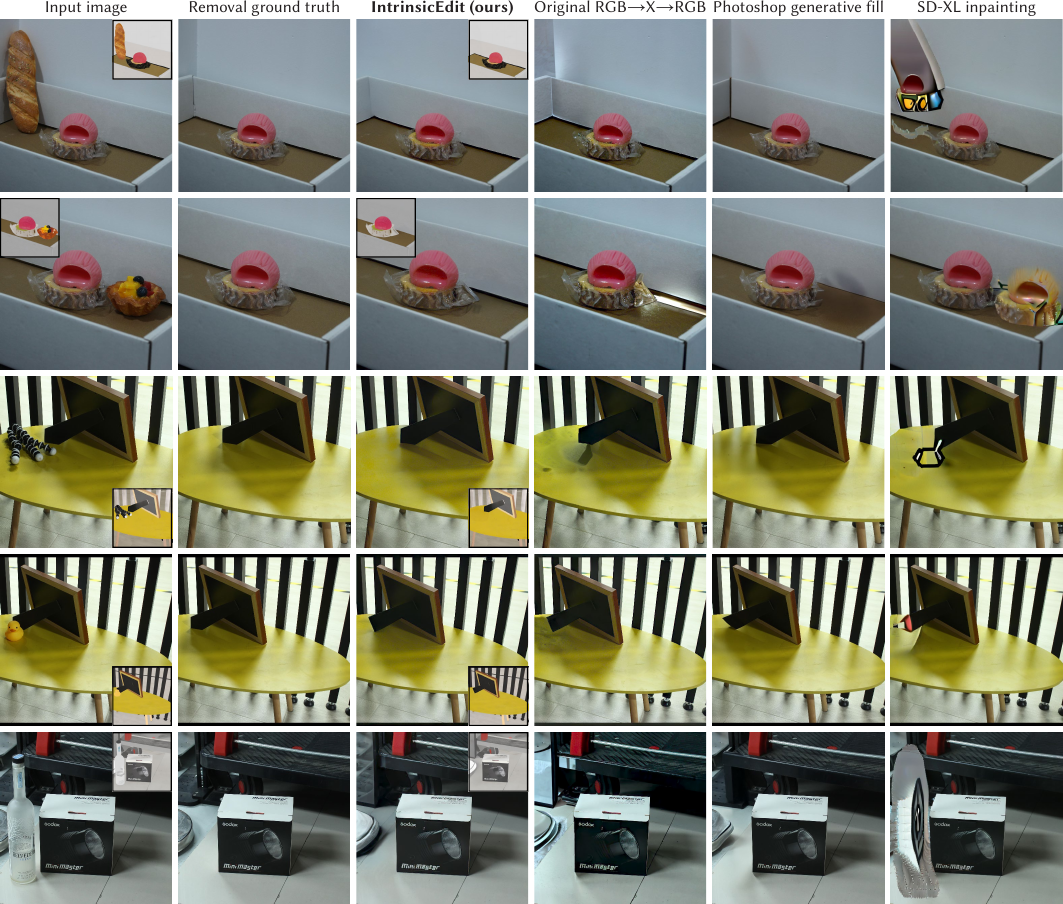}%
    \vspace{-2mm}
    \caption{
        \textbf{Real-world object removal A.}
        We include more results for the quantitative evaluation in the main paper of object removal on a real dataset.
    }
\end{figure*}


\begin{figure*}%
    \includegraphics{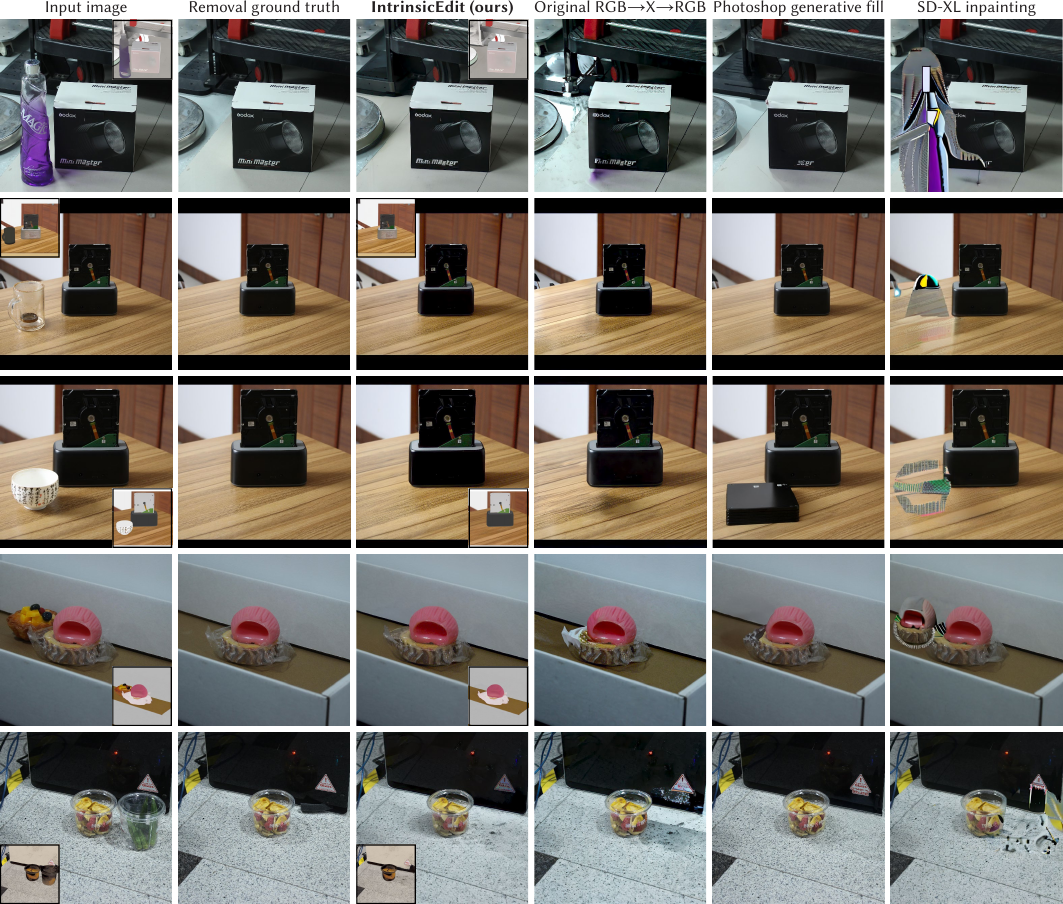}%
    \vspace{-2mm}
    \caption{
        \textbf{Real-world object removal B.}
        We include more results for the quantitative evaluation in the main paper of object removal on a real dataset.
    }
\end{figure*}


\begin{figure*}%
    \includegraphics{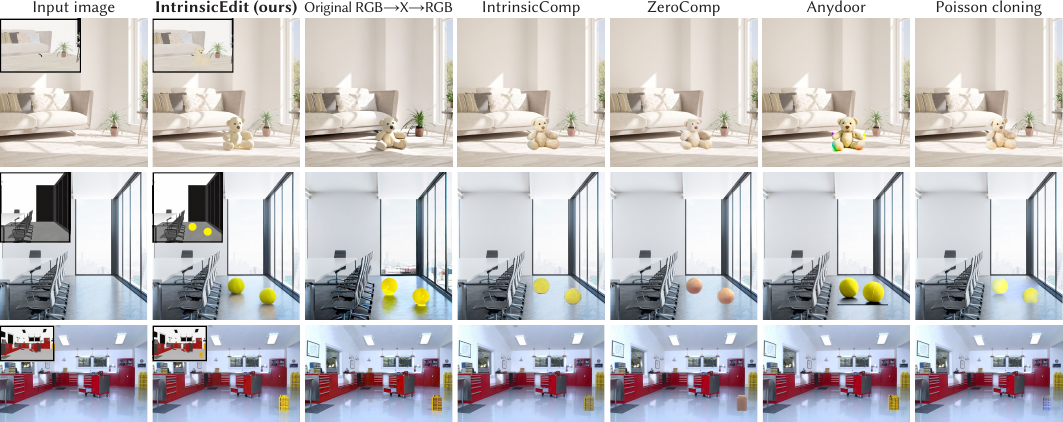}\\[0.65mm]%
    \includegraphics{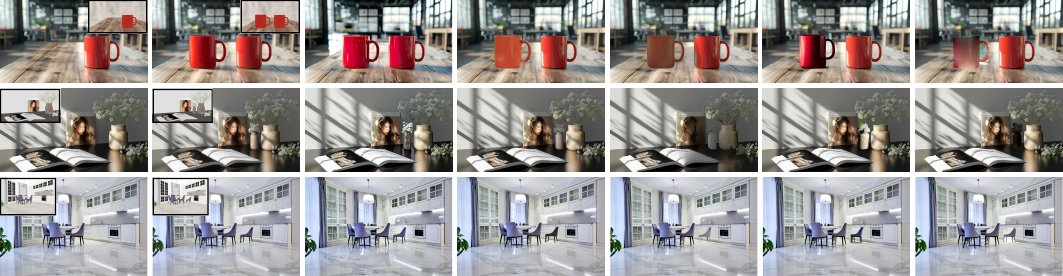}%
    \vspace{-2mm}
    \caption{
        \textbf{Object insertion A.}
        We compare against original \rgbxrgb \cite{zeng2024rgb} and existing object-insertion and intrinsic-based methods: IntrinsicComp \cite{careaga2023intrinsic}, ZeroComp \cite{zhang2024zerocomp}, Anydoor \cite{chen2023anydoor}, and Poisson cloning \cite{perez2003poisson}. For intrinsic-based methods, we insert the object into the albedo channel. Despite not being specialized for this task, our approach better harmonizes the inserted object with the rest of the scene.
    }
\end{figure*}


\begin{figure*}%
    \includegraphics{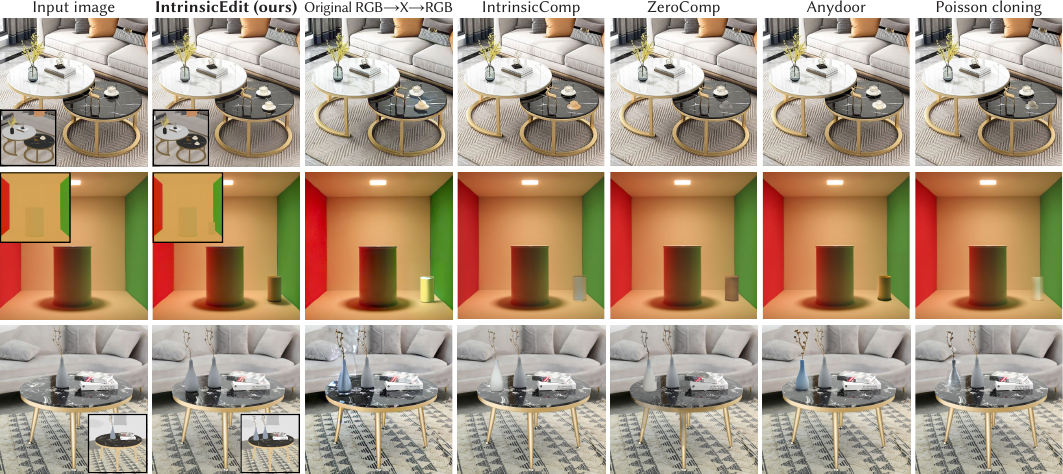}\\[0.65mm]%
    \includegraphics{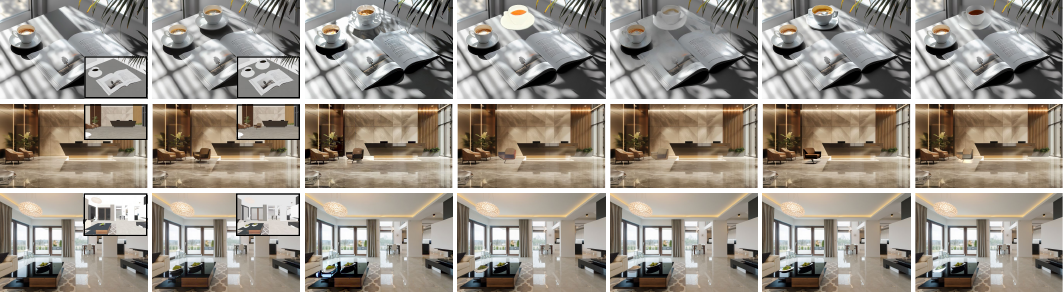}%
    \vspace{-2mm}
    \caption{
        \textbf{Object insertion B.}
        We compare against original \rgbxrgb \cite{zeng2024rgb} and existing object-insertion and intrinsic-based methods: IntrinsicComp \cite{careaga2023intrinsic}, ZeroComp \cite{zhang2024zerocomp}, Anydoor \cite{chen2023anydoor}, and Poisson cloning \cite{perez2003poisson}. For intrinsic-based methods, we insert the object into the albedo channel. Despite not being specialized for this task, our approach better harmonizes the inserted object with the rest of the scene.
    }
\end{figure*}


\begin{figure*}%
    \includegraphics{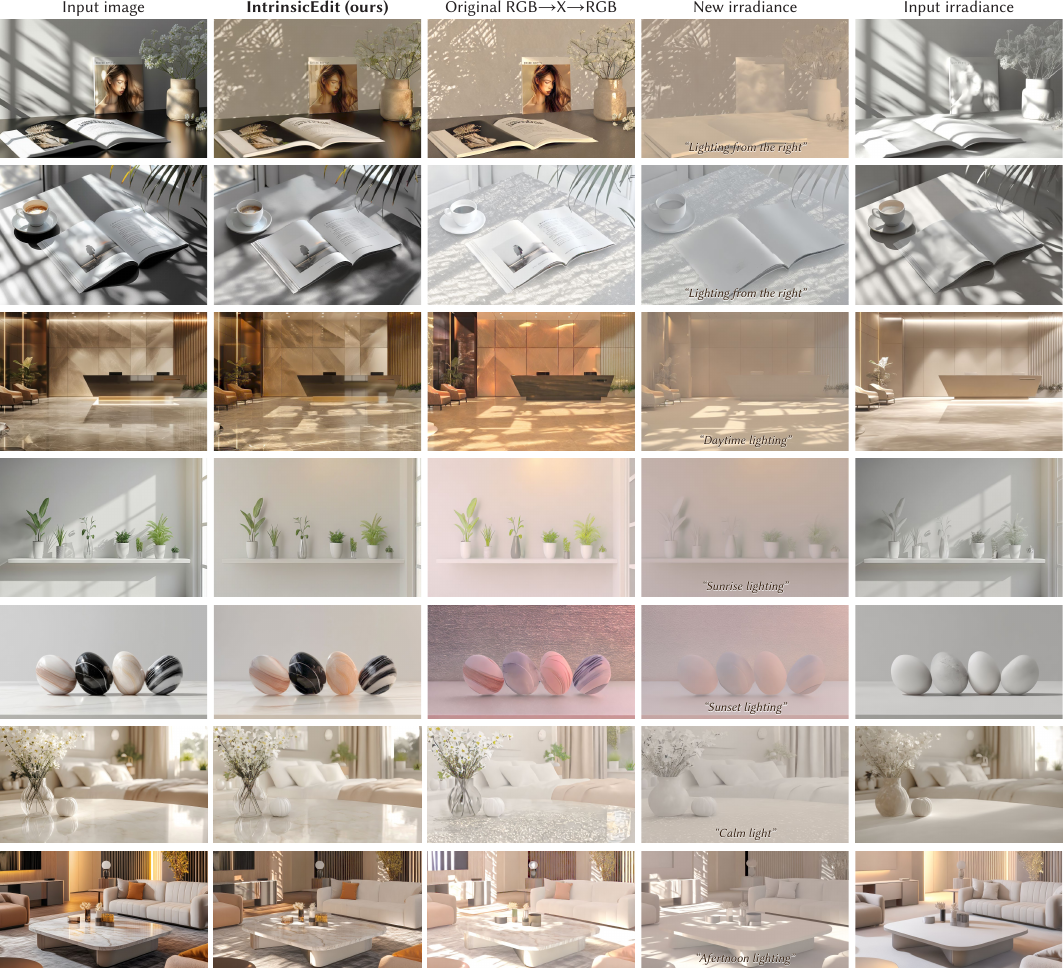}%
    \vspace{-2mm}
    \caption{
        \textbf{Relighting A.}
        We compare against original \rgbxrgb \cite{zeng2024rgb} relighting by changing the input irradiance. Our relighting handles the new lighting condition more naturally and better preserves the identity of the scene content.
    }
\end{figure*}


\begin{figure*}%
    \includegraphics{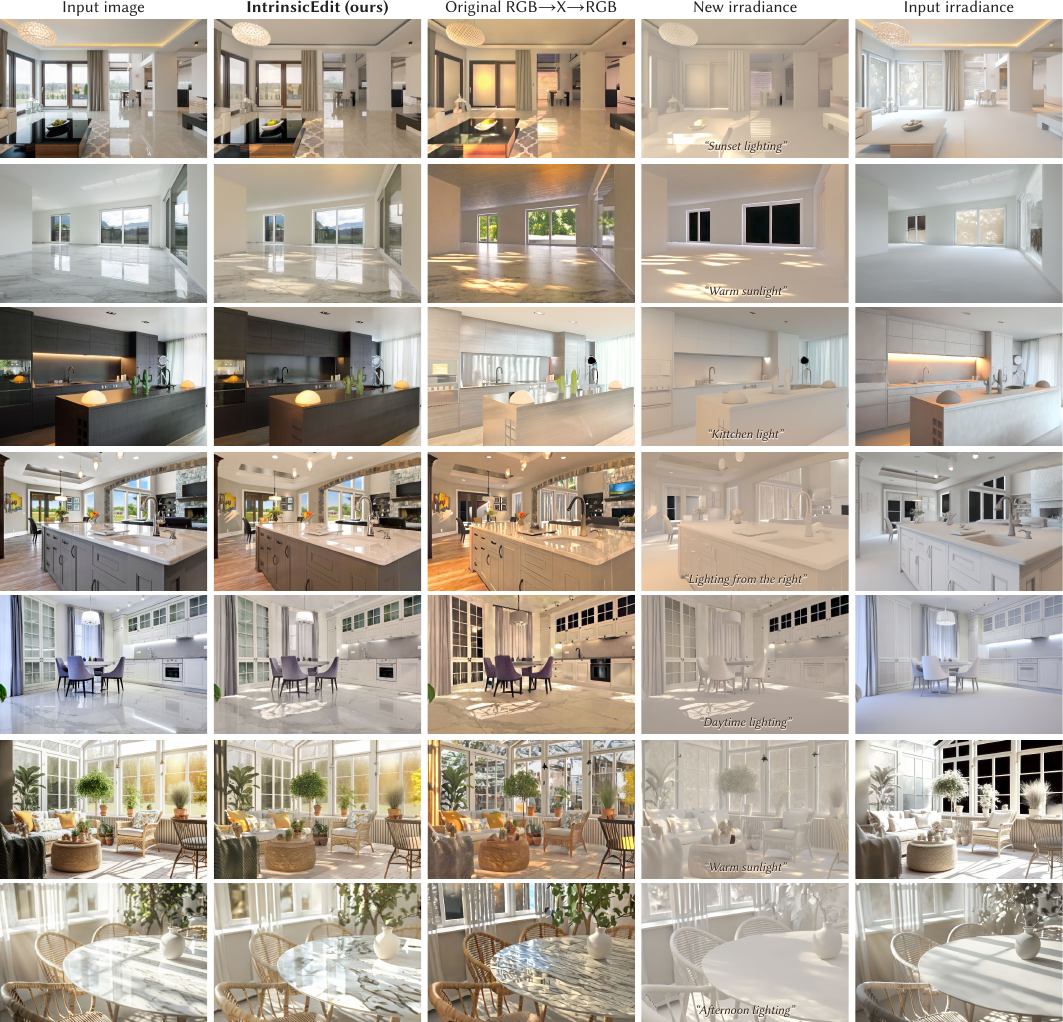}%
    \vspace{-2mm}
    \caption{
        \textbf{Relighting B.}
        We compare against original \rgbxrgb \cite{zeng2024rgb} relighting by changing the input irradiance. Our relighting handles the new lighting condition more naturally and better preserves the identity of the scene content.
    }
\end{figure*}


\bibliographystyle{ACM-Reference-Format}
\bibliography{references}
